\documentclass[aps,prc,twocolumn,groupedaddress]{revtex4-2}
\usepackage{amsmath,amssymb,multirow,graphicx,color}
\def\Vec#1{\boldsymbol{#1}}
\def\alignedequation#1{\begin{equation}\begin{aligned}#1\end{aligned}\end{equation}}

\begin{document}

%\preprint{}

\title{
 Symmetry classification of uniform states
%Correspondence between 
in
spin-2 Bose-Einstein condensates\\ 
and %Ginzburg-Landau theory for 
neutron $^3P_2$ superfluids
}


\author{Michikazu Kobayashi}
%\thanks{}
\affiliation{School of Environmental Science and Engineering, Kochi University of Technology, Miyanoguchi 185, Tosayamada, Kami, Kochi 782-8502, Japan}

\author{Muneto Nitta}
\affiliation{Department of Physics, and Research and Education Center for Natural Sciences, Keio University, Hiyoshi 4-1-1, Yokohama, Kanagawa 223-8521, Japan}


\date{\today}

\begin{abstract}
We clarify 
a relation between 
the Gross-Pitaevskii energy functional for 
spin-2 spinor Bose-Einstein condensates and 
the Ginzburg-Landau theory for neutron $^3P_2$ superfluidis (spin-triplet $P$-wave pairing with total angular momentum two).
We then classify all uniform states with nontrivial unbroken symmetries, with the help of geometric invariant theory.
\end{abstract}

%\keywords{}

\maketitle

\section{Introduction}

Condensations can have spin and angular momentum 
degrees of freedom. 
Superfluids $^3$He 
having both spin-triplet 
($S=1$) and $P$-wave ($L=1$) parings 
are such an example confirmed in nature \cite{vollhardt2013superfluid,volovik}. 
Among such condensations 
total-spin two ($J=2$) condensates consist of 
$2J+1=5$ complex degrees of freedom, 
and the
Ginzburg-Landau (GL) theory for
$J=2$ condensates was first studied 
by 
Mermin \cite{Mermin:1974zz}
in which possible ground states were classified. 
Nematic, cyclic, and ferromagnetic phases 
are typical phases. 
In nature, at least two examples of such condensations are known.
One is spin-2 ($S=2$) 
spinor Bose-Einstein condensations (BECs), 
which were realized in laboratory experiments.
The other is $^3P_2$ neutron superfluids consisting of 
neutron Cooper pairs 
of spin-triplet and $P$-wave paring  
with the total angular momentum $J=2$, 
relevant for neutron star cores.

Spin-2 spinor BECs can be 
theoretically 
well described by
the Gross-Pitaevskii (GP) equations 
and GP energy functional; 
see Ref.~\cite{2010arXiv1001.2072K} as a review.
There are typically three phases, 
cyclic, nematic, and ferromagnetic phases, 
and many interesting physics can be expected 
in each phase. 
Experimentally,  
spin-2 BECs are realized 
by $^{87}$Rb atoms, and the phase is around the boundary between 
cyclic phase and ferromagnetic phase
\cite{Schmaljohann:2004,Chang:2004,Kuwamoto:2004,Widera:2006,Tojo:2008,Tojo:2009}. 

In the cyclic phase,  
the order parameter manifold (OPM) 
parametrized by Nambu-Goldstone (NG) modes 
associated with spontaneous symmetry breaking 
is 
$[U(1) \times SO(3)]/T$ with a tetrahedral group $T$ \footnote{
The cyclic phase is also discussed in a superconductor \cite{Mizushima:2017pma}.
}.
Due to the nontrivial first homotopy group 
$\pi_1 \simeq {\mathbb Z} \times_h T^\ast$ 
($\times_h$ is a product defined in Ref.~\cite{Kobayashi:2011xb},
and $T^\ast$ is the universal covering group of $T$), 
1/3 quantized non-Abelian vortices are allowed 
\cite{Semenoff:2006vv,Kobayashi:2008pk,Mawson:2019}.
Two vortices belonging to noncommutative elements 
of $\pi_1$ do not reconnect in collisions, 
instead creating a rung vortex connecting them
\cite{Kobayashi:2008pk}. 
3D skyrmions based on the third homotopy group 
$\pi_3$ are allowed in the cyclic phase 
\cite{Tiurev:2017xgn}.

The nematic phase consists of three subphases 
continuously degenerated with each other:
uniaxial nematic (UN), $D_2$ biaxial nematic ($D_2$BN), 
and $D_4$ biaxial nematic ($D_4$BN) phases,
where $D_n$ is a dihedral group of order $n$.
The OPMs  
 are 
$U(1) \times SO(3)/O(2) \simeq S^1 \times {\mathbb R}P^2$, 
$U(1) \times SO(3)/D_2$ and $[U(1) \times SO(3)]/D_4$ of dimensions three, four, and four, respectively.
These are connected by a parameter of continuous degeneracy 
\cite{Uchino:2010},
which can be 
regarded as a quasi-NG mode associated with 
symmetry breaking of an enhanced symmetry 
\cite{Uchino:2010pf},
and these OPMs are 
submanifolds of an extended OPM 
$[S^1 \times S^4]/{\mathbb Z}_2$ 
parametrized by both the NG and quasi-NG modes. 
This continuous degeneracy can be lifted by 
quantum corrections \cite{Uchino:2010,Uchino:2010pf}.
In the nematic phase,  
1/2 quantized non-Abelian vortices are allowed 
\cite{Borgh:2016cco}.
As in the cyclic phase, 
3D skyrmions are also allowed in 
the BN phase \cite{Tiurev:2017xgn}.
 
%The topological defects on the boundary \cite{2019arXiv190702216C}.

On the other hand,
the neutron $^{3}P_{2}$ superfluids are relevant in the core of neutron stars~\cite{Tabakin:1968zz,Hoffberg:1970vqj,Tamagaki1970,Hoffberg:1970vqj,Takatsuka1971,Takatsuka1972,Fujita1972,Richardson:1972xn,Amundsen:1984qc,Takatsuka:1992ga,Baldo:1992kzz,Elgaroy:1996hp,Khodel:1998hn,Baldo:1998ca,Khodel:2000qw,Zverev:2003ak,Chamel:2008ca,Maurizio:2014qsa,Bogner:2009bt,Srinivas:2016kir,Haskell:2017lkl}.
See Refs.~\cite{Chamel:2008ca,Chamel2017,Haskell:2017lkl,Sedrakian:2018ydt,Graber:2016imq,Andersson2021}
as a  review from 
more general perspectives of 
superfluidity 
and superconductivity in neutron stars,
including neutron $^{3}P_{2}$ superfluids. 
The existence of
$^{3}P_{2}$ superfluids 
was indicated from astrophysical 
observations of 
the rapid cooling of a neutron star in Cassiopeia A~\cite{Heinke:2010cr,Shternin2011,Page:2010aw}. 
The GL theory as a bosonic effective theory around the transition point from the normal phase to the superfluid was established previously~\cite{Fujita1972,Richardson:1972xn,Sauls:1978lna,Muzikar:1980as,Sauls:1982ie,Vulovic:1984kc,Masuda:2015jka,Yasui:2018tcr,Yasui:2019tgc,Yasui:2019unp,
Yasui:2020xqb}. 
The ground state in the weak-coupling limit
was determined to be the nematic phase \cite{Sauls:1978lna}.
In the GL free energy up to the fourth order, 
there is a continuous degeneracy among UN, $D_2$BN, and $D_4$BN phases as the case of spin-2 BECs 
while a coupling to the magnetic field 
lifts the degeneracy, picking up  the $D_4$BN state
as the ground state \cite{Sauls:1978lna}.
In the absence of a magnetic field, 
the next-leading order of the GL expansion, namely,
the sixth order, lifts the degeneracy 
picking the UN phase as a possible ground state 
\cite{Sauls:1978lna}.
In the presence of the magnetic field, 
either UN, $D_2$BN, or $D_4$BN states can be the ground states 
depending on the strength of  the magnetic field \cite{Masuda:2015jka}.
However, at this order, the ground states are not global minima 
but just local minima because of the instability caused by the sixth order terms at large values of the condensates.
Thus, the GL free energy 
expanded up to the eighth order is needed to determine 
the unique ground states~\cite{Yasui:2019unp}. 
In the GL theory, 
 bosonic excitations as collective modes 
 in the $^3P_2$ superfluids were discussed 
\cite{Bedaque:2003wj,Leinson:2011wf,Leinson:2012pn,Leinson:2013si,Bedaque:2012bs,Bedaque:2013rya,Bedaque:2013fja,Bedaque:2014zta,Leinson:2009nu,Leinson:2010yf,Leinson:2010pk,Leinson:2010ru,Leinson:2011jr},
which are relevant for cooling process of neutron stars. 
Several kinds of vortices were discussed 
within the GL theory, 
such as 
integer vortices~\cite{Muzikar:1980as,Sauls:1982ie,Fujita1972,Masuda:2015jka,Chatterjee:2016gpm},
half-quantized non-Abelian vortices~\cite{Masuda:2016vak}, and 
coreless vortices \cite{Leinson:2020xjz}. 
Other topological defects such as
domain walls \cite{Yasui:2019vci} 
and 
topological defects on the boundary (called boojums) of $^3P_2$ superfluids~\cite{Yasui:2019pgb} were also found.

The Bogoliubov-de Gennes (BdG) equation 
offers a framework beyond the GL theory 
to deal with fermion degrees of freedom.  
The phase diagram in the plane of 
the temperature and magnetic field was determined 
in the BdG equation \cite{Mizushima:2016fbn}.
A tricritical point connecting 
first and second order phase transitions
 between $D_4$ and $D_2$ BN phases 
was found
\cite{Mizushima:2016fbn,Mizushima:2019spl}.
A topological superfluidity was also found 
\cite{Mizushima:2016fbn}, yielding
topologically protected 
Majorana zero modes on the boundary of $^3P_2$ superfluids  
\cite{Mizushima:2016fbn} 
and 
an integer vortex core 
\cite{Masaki:2019rsz}
as well as half-quantized vortex core 
\cite{Masaki:2021hmk}.

Thus far,
the so-called quasiclassical approximation linearizing dispersions around the Fermi surface was used to obtain nematic ground states 
in both the GL theory and BdG equation. 
Without the quasiclassical approximation, 
novel ground states were found  
around the critical temperature 
in the presence of strong magnetic fields: 
a ferromagnetic state 
and magnetized BN states (also called broken axisymmetric 
states in spin-2 BECs \cite{2010arXiv1001.2072K})
interpolating the ferromagnetic state 
and nematic state 
\cite{Mizushima:2021qrz}.

On the other hand, 
spin-2 BECs are closely related as one of candidates by astrophysical laboratories for neutron stars \cite{Graber:2016imq,Andersson2021}.
The low-energy theories such as GP theory for spinor BECs and GL theory for neutron superfluids give an important viewpoint to connect these systems.
However, these two theories
have been explored independently: 
condensates of spin-2 BECs are usually described 
by five component condensates $\psi$ 
and a $3 \times 3$ traceless symmetric tensor $A$  
are used for $^3P_2$ superfluids
in the literature,   
thus looking different at first glance.
However, since they are both $J=2$ condensates,  
they can be mapped to each other. 
The GP and GL energy functionals are quite similar
 after rewriting them, 
consequently admitting 
similar objects such as 
half-quantized non-Abelian vortices, 
\cite{Borgh:2016cco} and \cite{Masuda:2016vak}, 
respectively.
Nevertheless they are not exactly the same.
Apparent differences are, for instance, the following: 
(1) the GP theory contains up to the fourth order of the condensates, while the GL theory is basically expansion of the 
condensates, thus containing infinitely higher orders; 
(2) gradient terms are different: 
two tensor indices of the condensates of the spin-2 spinor BECs 
are both spins 
which are internal degrees of freedom, 
while one of two indices of the condensates of $^3P_2$ superfluids represents the angular momentum,
which can be contracted with spatial derivatives, 
thus admitting more gradient terms; and
(3) a magnetic field interacts with the condensates in different ways. 
These differences bring differences beyond topology 
such as stability and dynamics. 

The purpose of this paper is 
to clarify a relation between 
spin-2 spinor BECs and $^3P_2$ superfluids.
To this end, it turns out to be useful to use 
the geometric invariant theory 
(see, e.~g.~Refs.~\cite{Abud:1981tf,Abud:1983id,Nitta:1998qp}); 
since there are five 
condensations, the configuration space is
${\mathbb C}^5$.
The symmetry of the system 
is $U(1) \times SO(3)$, and then 
inequivalent configurations 
can be expressed by 
the ``moduli space of vacua''
${\mathbb C}^5/[U(1) \times SO(3)]$, 
parametrized by six $U(1) \times SO(3)$ invariants.
The symmetry classification of spin-2 spinor BECs have been 
investigated in terms of three invariants (among the totally six invariants) 
manifestly appearing in the low-energy Hamiltonian
\cite{Barnett:PRL2006,2010arXiv1001.2072K,Uchino:2010,Uchino:2010pf,Kawaguchi:PRA2011,Lian:PRA2012,
Stamper-Kurn:APS2013,Takahashi:2015}.
Here, we introduce remaining three invariants absent in the low-energy Hamiltonian of spin-2 BECs.
With a help of all these six invariants, 
we rewrite 
the energy functional of spin-2 BECs 
in terms of a traceless symmetric tensor, 
and that of $^3P_2$ superfluids 
in terms of five component condensates.
We find that in order to rewrite the latter in terms of the invariants, 
all six invariants are needed. 
We further classify all possible uniform states 
with nontrivial unbroken symmetries 
as candidates of ground states or metastable states, 
and clarify how they are connected in the moduli space.

This paper is organized as follows.
In Sec.~\ref{sec:formulation}, we introduce the standard 
formulations of 
 spin-2 BECs and $^3P_2$ superfluids. 
In Sec.~\ref{sec:relation}, we specify a relation between them, 
and in Sec.~\ref{sec:uniform} 
uniform states with nontrivial unbroken symmetries 
are classified 
as candidates of the ground states or metastable states. 
Section~\ref{sec:3P2} describes $^3P_2$ superfluids, and
Sec.~\ref{sec:summary} is devoted to a summary 
and discussion.

%\newpage

%%%%%%%%%%%%%%%%%%%%%%%%%%%%%%%%%%%%%%%%%

%%%%%%%%%%%%%%%%%%%%%%%%%%%%%%%%%%%%%%%%%

\section{Formulation}\label{sec:formulation}

In this section, we introduce the standard formulations 
of the GP energy functional of 
spin-2 BECs 
in terms of five component condensates, 
and the GL theory for $^3P_2$ superfluids 
in terms of a $3 \times 3$ traceless symmetric tensor.

%%%%%%%%
\subsection{Spin-2 spinor Bose-Einstein condensates}

The effective low-energy Hamiltonian density $h$ of spin-2 spinor BECs can be written as
\alignedequation{
h &= h_0 + h_{\rm qz} + h_{\rm int}, \\
h_0 &= \frac{\hbar^2}{2 m_{\rm b}} \Vec{j}^\dagger \Vec{j}, \\
h_{\rm qz} &= q \sum_{i,j=x,y,z} B_i \left( \psi^\dagger \hat{S}_i \hat{S}_j \psi \right) B_j, \\
h_{\rm int} &= \sum_{S = 0}^4 \frac{g_S}{2} \sum_{M = - S}^S \sum_{m_1, \cdots, m_4 = -2}^2 C_{2 m_1,2 m_2}^{S M} C^{S M}_{2 m_3,2 m_4} \\
&\times \psi_{m_1}^\ast \psi_{m_2}^\ast \psi_{m_3} \psi_{m_4},
\label{eq:spinor-Hamiltonian-original}
}
where $\psi = \begin{pmatrix}\psi_2 & \psi_1 & \psi_0 & \psi_{-1} & \psi_{-2} \end{pmatrix}^T$ is the five-component
condensate order parameters of a spin-2 BEC, $\Vec{j} = - i \nabla \psi$ is the current density, $m_{\rm b}$ is the mass of bosons, and $\hat{S}_i$ ($i = x,y,z$) are $5 \times 5$ spin-2 matrices.
In the quadratic Zeeman energy part $h_{\rm qz}$, $q = (g \mu_{B})^2 / E_{\rm hf}$ is the coefficient of the quadratic Zeeman energy under the magnetic field $\Vec{B}$ with the Land\'e $g$-factor $g$, the Bohr magneton $\mu_B$, and the hyperfine energy splitting $E_{\rm hf}$. 
In the interaction part $h_{\rm int}$, $g_{S}$ is the coupling constant in the total spin $S$ channel, and $C_{s_1 m_1,s_2 m_2}^{S M}$ are the Clebsch-Gordan coefficients.

The interaction part of the Hamiltonian density $h_{\rm int}$ can be rewritten as
\alignedequation{
& h_{\rm int} = \frac{1}{2} \left( c_0 \rho^2 + c_1 \Vec{S}^2 + c_2 |\Psi_{20}|^2 \right), \\
& \rho = \psi^\dagger \psi, \quad \Vec{S} = \psi^\dagger \hat{\Vec{S}} \psi, \\
& \Psi_{20} = \psi^T \hat{T} \psi = \sqrt{5} \sum_{m_1, m_2 = -2}^2 C^{00}_{2 m_1,2 m_2} \psi_{m_1} \psi_{m_2},
\label{eq:spinor-Hamiltonian}
}
where the coupling constants are defined by $c_0 = (4 g_2 + 3 g_4) / 7$, $c_1 = (g_4 - g_2) / 7$, $c_2 = (7 g_0 - 10 g_2 + 3 g_4) / 35$.
$\rho$, $\Vec{S}$, and $\Psi_{20}$ are known as the density, the spin density, and the singlet-pair amplitude with the time reversal operator $\hat{T}$ defined as $(\hat{T} \psi)_m = (-1)^m \psi_{-m} \equiv \varphi_m$. 
Note that $\rho^2$,  
$\Vec{S}^2$ and $|\Psi_{20}|^2$ are typical $U(1) \times SO(3)$ invariants.

In the absence (presence) of the magnetic field $B = 0$ ($B \neq 0$), the energy density is invariant under the transformation $\mathcal{G}_{\rm b}(\varphi,\Vec{n},\theta) \psi = e^{i \varphi} e^{- i \hat{\Vec{S}} \cdot \Vec{n} \theta} \psi$ ($\mathcal{G}_{\rm b}(\varphi,\Vec{B}/|\Vec{B}|,\theta) \psi =e^{i \varphi} e^{- i \hat{\Vec{S}} \cdot (\Vec{B}/|\Vec{B}|) \theta} \psi$), giving the symmetry $G_{\rm b} = U(1) \times SO(3)$ [$G_{\rm b} = U(1) \times SO(2)$]
of the energy density.

\subsection{Ginzburg-Landau free energy for 
$^3P_2$ neutron superfluids}
Here we briefly review 
the GL theory for $^3P_2$ superfluids 
in terms of a traceless symmetric tensor.
The original partition function of nonrelativistic two-spinor field $\varphi(t,\Vec{x}) = \begin{pmatrix} \varphi_\uparrow & \varphi_\downarrow \end{pmatrix}^T$
for a neutron is
\begin{align}
Z = \int \mathcal{D}\varphi\: \mathcal{D}\varphi^\ast \mathcal{D}A\: \mathcal{D}A^\ast\: \exp\left( - \int d\tau\: d\Vec{x}\: \mathcal{L} \right),
\label{eq:partition_function}
\end{align}
where $\tau \in [0, 1 / T]$ is the imaginary time $i t$ with the temperature $T$.
$\mathcal{L}$ is the Lagrangian density with the imaginary time $\tau$ as
\alignedequation{
& \mathcal{L} = \mathcal{L}_0 + \mathcal{L}_{\rm int}, \\
& \mathcal{L}_0 = \varphi^\ast \left( \partial_\tau + \frac{\nabla^2}{2 m_{\rm n}} + \mu + \Vec{\mu}_{\rm n} \cdot \Vec{B} \right) \varphi, \\
& \mathcal{L}_{\rm int} = \sum_{i,j = x,y,z} \left( A^{\ast ab} T^{ab} + T^{\ast ab} A^{ab} + \frac{A^{\ast ab} A^{ab}}{G} \right),
\label{eq:neutron_Lagrangian}
}
where $\Vec{B}$ is the external magnetic field, and $\Vec{\mu}_{\rm n} = - \hbar \gamma_{\rm n} \Vec{\sigma} / 2$ is the magnetic momentum with the Pauli matrices $\Vec{\sigma}$ and the neutron gyromagnetic ratio $\gamma_{\rm n}$ in the noninteracting Lagrangian density $\mathcal{L}_0$.
In the interacting Lagrangian density $\mathcal{L}_{\rm int}$, $G$ is the renormalized coupling constant, $T^{ab}$ is the tensor operator for the repulsive interaction, and $A^{ab}$ is the auxiliary tenser field having symmetric and traceless form.

Near the transition temperature $T_{\rm c}$ for neutron $^3P_2$ superfluidity, the original partition function \eqref{eq:partition_function} gives the GL free-energy density $f$ as 
\cite{Yasui:2019unp}
%\cite{Yasui2019}
\alignedequation{
f &= K_0 \left(f_{202}^{(0)} +f_{202}^{(1)}\right) + \alpha f_{002} + \beta_0 f_{004} + \gamma_0 f_{006} \\
&+ \delta_0 f_{008} + \beta_2 f_{022} + \gamma_2 f_{024} \\
&+ \sum_{4 l + 2 m + n = 10} \mathcal{O}(\partial^l |\Vec{B}|^m A^n).
\label{eq:neutron_Free_energy}
}
Each term in the free-energy density $f$ can be written as
\begin{widetext}
\alignedequation{
f_{202}^{(0)} &= \sum_{i,j,k = x,y,z} \partial_i A_{jk} \partial_i A^\ast_{jk}, \quad
f_{202}^{(1)} = \sum_{i,j,k=x,y,z} \left( \partial_i A_{ji} \partial_k A_{jk}^\ast + \partial_i A_{jk} \partial_k A_{ji}^\ast \right), \\
f_{002} &= \left(\mathrm{tr}A^\ast A\right), \quad
f_{004} = \left(\mathrm{tr}A^\ast A\right)^2 - \left(\mathrm{tr}A^{\ast 2} A^2\right), \\
f_{006} &= - 24 \left(\mathrm{tr}A^\ast A\right)^3 + 36 \left(\mathrm{tr}A^\ast A\right) \left(\mathrm{tr}A^{\ast 2} A^2 \right) - 6 \left|\mathrm{tr}A^\ast A^2\right|^2 + \frac{10}{3} \left|\mathrm{tr}A^3\right|^2, \\
f_{008} &= 192 \left(\mathrm{tr}A^\ast A\right)^4 - 384 \left(\mathrm{tr}A^\ast A\right)^2 \left(\mathrm{tr}A^{\ast 2} A^2\right) - 64 \left(\mathrm{tr}A^\ast A\right) \left|\mathrm{tr}A^3\right|^2 + 144 \left(\mathrm{tr}A^{\ast 2} A^2\right)^2 + 192 \left(\mathrm{tr}A^\ast A\right) \left|\mathrm{tr}A^\ast A^2\right|^2 \\
&- 48 \left|\mathrm{tr}A^2\right|^2 \left(\mathrm{tr}A^{\ast 2} A^2\right) + 12 \left|\mathrm{tr}A^2\right|^4 - 96\: \mathrm{Re}\left[\left(\mathrm{tr}A^2\right) \left(\mathrm{tr}A^{\ast 2} A\right)^2\right], \\
f_{022} &= \Vec{B}^T \left(A^\ast A\right) \Vec{B}, \\
f_{024} &= \Vec{B}^2 \left\{ - 2 \left(\mathrm{tr}A^{\ast 2}\right) \left(\mathrm{tr}A^2\right) - 4 \left(\mathrm{tr}A^\ast A\right)^2 + 4 \left(\mathrm{tr}A^\ast A A^\ast A\right) + 8 \left(\mathrm{tr}A^{\ast 2} A^2\right)\right\} \\
  &+ 2\: \mathrm{Re}\left[\left( \mathrm{tr}A^2 \right) \left( \Vec{B}^T A^{\ast 2} \Vec{B} \right)\right] - 8 \left( \mathrm{tr}A^\ast A \right) \left( \Vec{B}^T A^\ast A \Vec{B} \right) + 4 \Vec{B}^T \left( \mathrm{Re}\left[A^\ast A^2 A^\ast\right] - 2 A^\ast A A^\ast A - 2 A^{\ast 2} A^2 \right) \Vec{B}.
\label{eq:neutron_terms}
}
\end{widetext}
The GL coefficients can be calculated in the weak coupling limit 
within the quasiclassical approximation as 
\cite{Yasui:2019unp}

\begin{equation}
\begin{alignedat}{2}
K_0 &= \frac{7 \zeta(3) N(0) p_{\rm F}^4}{240 \pi^2 m_{\rm n}^2 T^2},\quad & \alpha &= \frac{N(0) p_{\rm F}^2}{3} \log\frac{T}{T_{\rm c}},\\
\beta_0 &= \frac{7 \zeta(3) N(0) p_{\rm F}^4}{60 \pi^2 T^2},\quad & \gamma_0 &= \frac{31 \zeta(5) N(0) p_{\rm F}^6}{13440 \pi^4 T^4},\\
\delta_0 &= \frac{127 \zeta(7) N(0) p_{\rm F}^8}{387072 \pi^6 T^6},\quad & \beta_2 &= \frac{7 \zeta(3) N(0) p_{\rm F}^2 \gamma_{\rm n}^2}{48 (1 + F_0^a)^2 \pi^2 T^2},\\
\gamma_2 &= \frac{31 \zeta(5) N(0) p_{\rm F}^4 \gamma_{\rm n}^2}{3840 (1 + F_0^a)^2 \pi^4 T^4}, & &
\end{alignedat}
\end{equation}
with the neutron mass $m_{\rm n}$, the Fermi momentum $p_{\rm F}$, the state-number density $N(0) = m_{\rm n} p_{\rm F} / (2 \pi)^2$ at the Fermi surface, and the Landau parameter $F_0^a$ defined as the modification of the magnetic momentum $|\Vec{\mu}| = (\gamma_{\rm n} \hbar / 2) / (1 + F_0^a)$.
The GL free energy without 
 the quasiclassical approximation
 was also calculated in Ref.~\cite{Mizushima:2021qrz}.

One can consider the transformation with the spin-orbit locked symmetry
\begin{equation} 
\begin{aligned}
\bar{\mathcal{G}}_{\rm n}(\varphi, \Vec{n}, \theta) A = e^{i \varphi} R(\Vec{n},\theta) A R(\Vec{n},\theta)^T,
\end{aligned}
\label{eq:neutron_transformation}
\end{equation}
%$U(1)_{\rm B} \times SO(3)_{\rm S} \times SO(3)_{\rm L}$,
%\begin{equation} 
%\bar{\mathcal{G}}_{\rm n}(\varphi_{\rm B}, \Vec{n}_{\rm S},\Vec{n}_{\rm L}, \theta_{\rm S}, \theta_{\rm L}) A = e^{i \varphi_{\rm B}} R(\Vec{n}_{\rm S},\theta_{\rm S}) A R(\Vec{n}_{\rm L},\theta_{\rm L})^T,
%\end{equation}
where $e^{i \varphi}$ denotes the global $U(1)$ phase shift and $R(\Vec{n},\theta)$ denotes spin-orbit locked rotation with the rotation axes $\Vec{n}$ and rotation angles $\theta$.
The free-energy density is invariant under the transformation shown in Eq.~\eqref{eq:neutron_transformation} 
%of the spin-orbit locked symmetry  
%\begin{equation} 
%\bar{\mathcal{G}}_{\rm n}(\varphi_{\rm B}, \Vec{n}_{\rm S+L}) A = e^{i \varphi_{\rm B}} R(\Vec{n}_{\rm S+L},\theta_{\rm S+L}) A R(\Vec{n}_{\rm S+L},\theta_{\rm S+L})^T,
%\end{equation}
in the absence of the magnetic field $\Vec{B} = 0$, 
showing the symmetry $G_{\rm n} \simeq U(1) \times SO(3)$ for the free energy. 

\if0 %%
When $A$ is fixed, the transformation $\bar{\mathcal{G}}_{\rm n}(\varphi_{\rm B}, \Vec{n}_{\rm S},\Vec{n}_{\rm L}, \theta_{\rm S}, \theta_{\rm L}) A$ can be rewritten as $\mathcal{G}_{\rm n}(\varphi_{\rm B}, \Vec{n}_{{\rm S+L},A}, \theta) A = R(\Vec{n}_{{\rm S+L},A},\theta_{\rm S+L}) A R(\Vec{n}_{{\rm S+L},A},\theta_{\rm S+L})^T$ with an appropriate choice of $\Vec{n}_{{\rm S+L},A}$ and $\theta_{\rm S+L}$, giving the symmetry breaking from $G_{\rm n}$ to $G^\prime_{\rm n} \simeq U(1) \times SO(3)$.
\fi %%
In the presence of the magnetic field $\Vec{B} \neq 0$, the symmetry $G_{\rm n}$ %and $G^\prime_{\rm n}$ are changed as $G_{\rm n} \simeq U(1)_{\rm B} \times SO(2)_{\rm S} \times SO(2)_{\rm L}$ and  $G^\prime_{\rm n} \simeq U(1)_{\rm B} \times SO(2)_{\rm S+L}$, where the rotational axis $\Vec{n}_{\rm S+L}$ is fix to be parallel to $\Vec{B}$. 
is explicitly broken by the background to  
%$G_{\rm n} \simeq U(1)_{\rm B} \times SO(2)_{\rm S} \times SO(2)_{\rm L}$ and  
$G_{\rm n} \simeq U(1) \times SO(2)$, where the rotational axis $\Vec{n}$ is fix to be parallel to $\Vec{B}$.

\section{Relation between neutron superfluidity and spin-2 spinor condensates}\label{sec:relation}

In this section, we rewrite the free-energy density $f$ for neutron $^3P_2$ superfluids and the energy density $h$ for spin-2 spinor condensates in terms of the condensate order parameter $\psi$ and the tensor field $A$, respectively, by introducing a map between them.

\subsection{Spin-2 spinor BECs in terms of a traceless 
symmetric tensor $A$}
We begin with considering the spherical harmonic form $Y_\psi$ for the condensate wave function $\psi$ in Eq. \eqref{eq:spinor-Hamiltonian} as
\begin{align}
Y_\psi = \sum_{m=-2}^2 Y_{2\,m}(\Vec{n}) \psi_m,
\label{eq:spherical_form}
\end{align}
where $Y_{2m}(\Vec{n})$ is the rank-2 spherical harmonic
functions
\alignedequation{
Y_{2\,\pm2} &= \frac{1}{4} \sqrt{\frac{15}{2 \pi}} (n_x \pm i n_y)^2, \\
Y_{2\,\pm1} &= \mp \frac{1}{2} \sqrt{\frac{15}{2 \pi}} n_z (n_x \pm i n_y), \\
Y_{2\,0} &= \frac{1}{4} \sqrt{\frac{5}{\pi}} (2 n_z^2 - n_x^2 - n_y^2),
\label{eq:spherical_harmonic}
}
with the unit vector $\Vec{n}$.
The spherical harmonic form $Y_\psi$ in Eq. \eqref{eq:spherical_form} can be rewritten as
%\begin{align}
%Y_\psi = \frac{1}{4} \sqrt{\frac{15}{2 \pi}} \begin{pmatrix} n_x & n_y & n_z \end{pmatrix}
%A
%\begin{pmatrix} n_x \\ n_y \\ n_z \end{pmatrix},
%\label{eq:translation}
%\end{align}
where $A$ is a $3 \times 3$ complex matrix given by
\alignedequation{
& [A]_{11} = \frac{\sqrt{3}}{2} (\psi_2 + \psi_{-2}) - \frac{1}{\sqrt{2}} \psi_0, \\
& [A]_{12} = [A]_{21} = \frac{\sqrt{3} i}{2} (\psi_2 - \psi_{-2}), \\
& [A]_{13} = [A]_{31} = - \frac{\sqrt{3}}{2} (\psi_1 - \psi_{-1}), \\
& [A]_{22} = - \frac{\sqrt{3}}{2} (\psi_2 + \psi_{-2}) - \frac{1}{\sqrt{2}} \psi_0, \\
& [A]_{23} = [A]_{32} = - \frac{\sqrt{3} i}{2} (\psi_1 + \psi_{-1}), \\
& [A]_{33} = \sqrt{2} \psi_0.
\label{eq:translation}
}
Because $A$ is the traceless and symmetric having the same property with the order parameter $A$ in
Eq.~\eqref{eq:neutron_Free_energy}, we can regard
Eq.~\eqref{eq:translation} as a map between the order parameters  of neutron $^3P_2$ superfluids and spin-2 spinor condensates.
By using Eq.~\eqref{eq:translation}, the density $\rho$, the spin density $\Vec{S}^2$, and the singlet-pair amplitude $\Psi_{20}$ can be rewritten as
\alignedequation{
\rho &= \frac{1}{3} \left( \mathrm{tr} A^\ast A \right), \quad
\Psi_{20} = \frac{1}{3} \left(\mathrm{tr} A^2 \right), \\
\Vec{S}^2 &= \frac{4}{9} \left( \mathrm{tr} A^\ast A \right)^2 + \frac{2}{9} \left|\mathrm{tr} A^2\right|^2 - \frac{4}{3} \left(\mathrm{tr}A^{\ast 2} A^2\right).
}
The Hamiltonian density $h$ in Eq.~\eqref{eq:spinor-Hamiltonian} for the spinor condensates can be rewritten as
\alignedequation{
& h_0 = \frac{\hbar^2}{6 m_{\rm b}} \sum_{i,j,k=x,y,z} \partial_i A_{jk} \partial_i A_{jk}^\ast, \\
& h_{\rm qz} = - 2 q \left\{ \Vec{B}^T \left( A^\ast A \right) \Vec{B} - \frac{2}{3} \left( \mathrm{tr} A^\ast A \right) \Vec{B}^2 \right\} \\
& h_{\rm int} = \frac{c_0 + 4 c_1}{18} \left( \mathrm{tr} A^\ast A \right)^2 - \frac{2 c_1}{3} \left( \mathrm{tr} A^{\ast2} A^2 \right) \\
&\phantom{h_{\rm int}} + \frac{2 c_1 + c_2}{18} \left( \mathrm{tr} A^{\ast2}  \right) \left( \mathrm{tr} A^2 \right)
}
in terms of the traceless symmetric tensor $A$.
Apart from apparent absence of higher order terms of $A$, 
one can observe that the gradient term $f_{202}^{(1)}$  
in Eq.~(\ref{eq:neutron_terms}) 
for $^3P_2$ superfluids is absent in this case.
At the fourth order, the last term in 
$h_{\rm int}$ is absent in Eq.~(\ref{eq:neutron_terms}).
As for the interaction with the magnetic field $\Vec{B}$, 
the second term in  $h_{\rm qz} $ is absent 
 in Eq.~(\ref{eq:neutron_terms}) at this order.

\subsection{$^3P_2$ superfluids in terms of five component condensates $\Psi$}
On the other hand, the free energy density $f$ in
Eq.~\eqref{eq:neutron_Free_energy} for neutron $^3P_2$ superfluids can be rewritten as
\alignedequation{
f_{202}^{(0)} &= 3 \Vec{j}^\dagger \cdot \Vec{j}, \\
f_{202}^{(1)} &= 4 \Vec{j}^\dagger \cdot \Vec{j} - \frac{i}{2} \Vec{j}^\dagger \cdot \hat{\Vec{S}} \times \Vec{j} - \left( \Vec{j}^\dagger \cdot \hat{\Vec{S}} \right) \left( \hat{\Vec{S}} \cdot \Vec{j} \right), \\
f_{002} &= 3 \rho, \quad
f_{004} = 6 \rho^2 + \frac{3}{4} \Vec{S}^2 - \frac{3}{2} |\Psi_{20}|^2, \\
f_{006} &= - 324 \rho^3 - 81 \rho \Vec{S}^2 + 162 \rho |\Psi_{20}|^2 \\
&+ 15 |\Psi_{30}|^2 - 27 |\Phi_{30}|^2, \\
f_{008} &= 6480 \rho^4 + 1944 \rho^2 \Vec{S}^2 - 5184 \rho^2 |\Psi_{20}|^2 \\
&- 864 \rho |\Psi_{30}|^2 + 2592 \rho |\Phi_{30}|^2 + 81 \Vec{S}^4 \\
&+ 648 |\Psi_{20}|^4 - 1296 \Gamma_{4} \\
f_{022} &= 2 \rho \Vec{B}^2 - \frac{1}{2} \psi^\dagger \hat{S}_{\Vec{B}} \hat{S}_{\Vec{B}} \psi, \\
f_{024} &= \left( -100 \rho^2 + 3 \Vec{S}^2 + 16 |\Psi_{20}|^2 \right) \Vec{B}^2 \\
&+ \left\{ 22 \rho \psi^\dagger \hat{S}_{\Vec{B}} \hat{S}_{\Vec{B}} \psi + \mathrm{Re}\left[ \Psi_{20}^\ast \psi^T \hat{S}_{\Vec{B}}^T \hat{T} \hat{S}_{\Vec{B}} \psi \right] \vphantom{\frac{5}{4}} \right. \\
&\phantom{+} + \left. \frac{5}{4} \Psi_{22}^\dagger \hat{S}_{\Vec{B}} \hat{S}_{\Vec{B}} \Psi_{22} + \frac{1}{2} \Phi_{22}^T \hat{S}_{\Vec{B}}^T \hat{T} \hat{S}_{\Vec{B}} \Phi_{22} \right\},
}
with $\hat{S}_{\Vec{B}} \equiv \hat{\Vec{S}} \cdot \Vec{B}$,
where $\Psi_{30}$ and $\Phi_{30}$ are 
$SO(3)$-invariants, called the singlet trio and the deformed single trio amplitudes, respectively, defined by
\alignedequation{
\Psi_{30}
&= \frac{\sqrt{2}}{3}\mathrm{tr} \left(A^3\right) \\
&= - \sqrt{\frac{35}{2}} \sum_{J=0}^4 \sum_{M = -J}^J \sum_{m_1, m_2, m_3 = -2}^2 \\
&\times C^{00}_{JM,2m_3} C^{JM}_{2m_1,2m_2} \psi_{m_1} \psi_{m_2} \psi_{m_3}, \\
\Phi_{30}
& = \frac{\sqrt{2}}{3} \mathrm{tr}\left(A^\ast A^2\right) \\
&= - \sqrt{\frac{35}{2}} \sum_{J=0}^4 \sum_{M = -J}^J \sum_{m_1, m_2, m_3 = -2}^2 \\
&\times C^{00}_{JM,2m_3} C^{JM}_{2m_1,2m_2} \psi_{m_1} \psi_{m_2} \varphi^\ast_{m_3},
}
respectively, and 
$\Gamma_4$ is an $U(1) \times SO(3)$ invariant, defined by
\alignedequation{
\!\!\!\Gamma_4
= \mathrm{Re}\left[ \Psi_{20} \Phi_{30}^{\ast 2} \right]
= \frac{2}{27} \mathrm{Re}\left[ \mathrm{tr}\left(A^2\right) \left\{ \mathrm{tr}\left(A^{\ast 2} A\right) \right\}^2 \right].
}
Here, $\Psi_{22}$ and $\Phi_{22}$ are the quintuplet (spin-2) pair and the deformed quintuplet pair amplitudes, respectively, defined by
\alignedequation{
& \left( \Psi_{22} \right)_M = \sqrt{14} \sum_{m_1, m_2 = -2}^2 C^{2 M}_{2 m_1 2 m_2} \psi_{m_1} \psi_{m_2}, \\
& \left( \Phi_{22} \right)_M = \sqrt{14} \sum_{m_1, m_2 = -2}^2 C^{2 M}_{2 m_1 2 m_2} \psi_{m_1} \varphi_{m_2}^\ast,
}
where $M$ runs from $-2$ to $2$ for the total spin 2 channel with the Clebsch-Gordan coefficient $C^{2 M}_{2 m_1 2 m_2}$.

%%%%%%%%%%%%%%%%%%%%%%%%%%%%
\section{Classifying uniform states} \label{sec:uniform}

In this section, we classify  the all possible uniform states 
with nontrivial unbroken symmetries 
for $f$ in Eq.~\eqref{eq:neutron_Free_energy} and $h$
in Eq.~\eqref{eq:spinor-Hamiltonian}, 
which are called strata in the geometric inviariant theory, 
as candidates 
of the ground states or metastable states, 
and determine the moduli space 
to which all the uniform states belong.

\subsection{The geometric structure}

When $\Vec{B} = 0$, uniform part of the free-energy density
$\alpha f_{002} + \beta_0 f_{004} + \gamma_0 f_{006} + \delta_0 f_{008}$ and the Hamiltonian density
$h_{\rm int}$ should include only terms which are invariant under the $U(1) \times SO(3)$
transformation $\mathcal{G}_{\rm n,b}$.
Since $A$ or $\psi$ belongs to 
${\mathbb C}^5$ and
its real dimensions (or degrees of freedom) are 10,
% with fixing $\mathrm{tr} A$ or $\rho$, respectively, 
there should be 
$10- \dim [U(1) \times SO(3)] = 6$ independent $U(1) \times SO(3)$ invariants  
composed of $A$ or $\psi$.
We can choose them as 
\alignedequation{
 \left\{ \begin{array}{c} \rho, \Vec{S}^2,  |\Psi_{20}|^2, \\[3pt]
 |\Psi_{30}|^2, |\Phi_{30}|^2,  \Gamma_4 \end{array} \right\}
 \in \frac{\mathbb{C}^5}{U(1) \times SO(3)} \equiv {\cal M}.
\label{eq:5inv}
}
All of other $U(1) \times SO(3)$ invariants such as $f_{00n\geq10}$ can be
written only with these six terms and are not independent.
This is a $U(1) \times SO(3)$ orbit space 
in which subspace with the same unbroken symmetries 
are called strata.

In the following, we exclude the case of $\rho = 0$ in which the $U(1) \times SO(3)$ symmetry is fully recovered.
Furthermore, we always fix the value of $\rho (> 0)$, 
because it is irrelevant to the symmetry structure of the
uniform states.
With fixing $\rho$ and dividing it by $U(1)$ first, we obtain the complex complex projective space
$\mathbb{C}P^4$ of the real dimension eight.
Dividing it by $SO(3)$, we have a five-dimensional space
\begin{equation}
 \left\{ \begin{array}{c} \Vec{S}^2,  |\Psi_{20}|^2, \\[3pt]
 |\Psi_{30}|^2,  |\Phi_{30}|^2,  \Gamma_4 \end{array} \right\} 
 \in \frac{\mathbb{C}P^4}{SO(3)} \simeq {\cal M}_{\rho=1} \subset {\cal M},
\end{equation}
parametrized by the five $U(1) \times SO(3)$ invariants instead of Eq.~\eqref{eq:5inv}.

\begin{figure}[htb]
\centering
\includegraphics[width=0.65\linewidth]{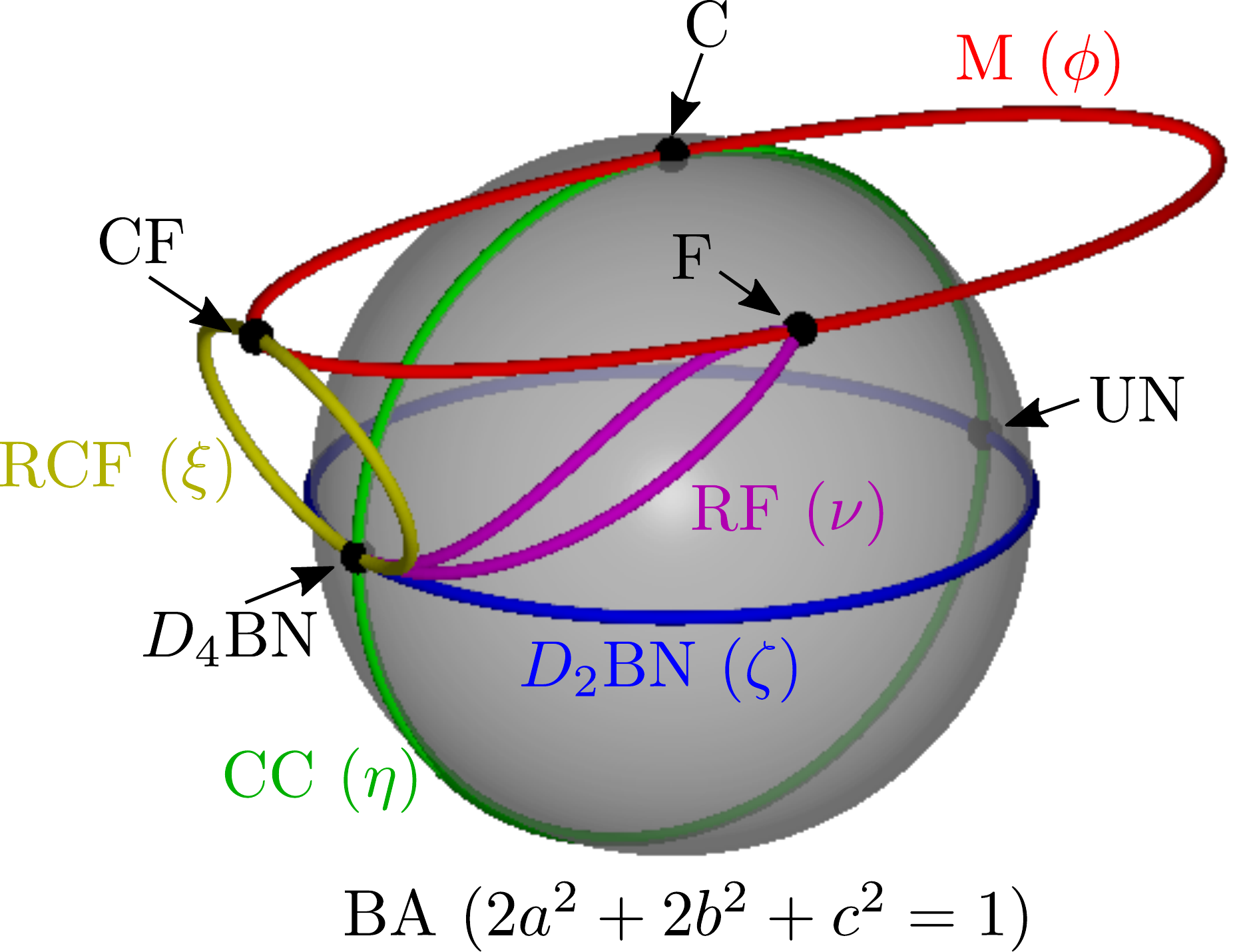}
\caption{
\label{fig:phase_image}
Schematic image for the topological structure of uniform states having the nontrivial internal symmetries $H_{\rm n,b}$ under $\Vec{B} = 0$.
}
\end{figure}
When the uniform state $A$ in $f$ or $\psi$ in $h$ is fixed,
the symmetries of the system are further broken from $G_{\rm n,b}$
down to subgroups $H_{\rm n,b}$, where $H_{\rm n,b}$ are the
symmetries of the uniform state $A$ and $\psi$
Depending on the nontrivial unbroken symmetries
$H_{\rm n,b}$, there can be nine (15) candidates for 
the ground (or metastable) states (or strata) 
with $\Vec{B} = 0$ ($\Vec{B} \neq 0$).

We define $\mathcal{S}$ ($\subset \mathcal{M}_{\rho = 1}$) by a space of all uniform states having the nontrivial unbroken symmetry $H_{\rm n,b}$ under $\Vec{B} = 0$
except for the trivial state $\rho = 0$ where the $U(1) \times SO(3)$ symmetry is fully recovered.
In Fig.~\ref{fig:phase_image}, we show a schematic image of $\mathcal{S}$, consisting of
%Before discussing each of the uniform states,
%we show in Fig.~\ref{fig:phase_image} a schematic image for the topological structure
%of all uniform states having the nontrivial unbroken symmetry $H_{\rm n,b}$ under $\Vec{B} = 0$
%except for the trivial state $\rho = 0$ where the $U(1) \times SO(3)$ symmetry is fully recovered. 
%The whole structure consists of
a gray $S^2$ surface, green, blue, and purple $S^1$ curves 
on the $S^2$ surface,
a red $S^1$ curve connected to the $S^2$ surface
at two points, 
and a yellow $S^1$ curve attached 
to the $S^2$ and the red curve at points.
The dimension for this space is $\dim[\mathcal{S}] = 2$.
This is a $U(1) \times SO(3)$ orbit space 
with nontrivial unbroken symmetries which are not 
$U(1) \times SO(3)$ or $1$. 
The $U(1) \times SO(3)$ orbits are fibered over $\mathcal{S}$ 
to recover $S^9$ of $A$ or $\psi$ with the constraint $\rho = 1$.
However,
because the total dimension of the space for states having the nontrivial symmetries is 
$\dim[ \{U(1) \times SO(3)\} \times \mathcal{S}] = 6$,
this space cannot cover $S^9$ space for $A$ or $\psi$ having nine dimensions.
All states having the trivial internal symmetry $H_{\rm n,b} \simeq 1$ constructs the
remaining five-dimensional space $(M_{\rho=1} - \mathcal{S})$.

Below we show all of the uniform states having the nontrivial 
unbroken symmetries $H_{\rm n,b}$ one by one.
%When referring the symmetry $H_{\rm n,b}$ for $\Vec{B} \neq 0$, we assume $\Vec{B} \parallel \hat{\Vec{z}}$.

\subsection{Ferromagnetic states}

The ferromagnetic (F) state $\psi_{\rm F}$ can be written as
\alignedequation{
\bar{\psi}_{\rm F} &= \begin{pmatrix} 1 & 0 & 0 & 0 & 0 \end{pmatrix}^T, \\
\bar{A}_{\rm F} &= \frac{\sqrt{3}}{2} \begin{pmatrix} 1 & i & 0 \\ i & -1 & 0 \\ 0 & 0 & 0 \end{pmatrix},
}
where the overbar denotes the normalization as $\bar{\psi} = \psi / \sqrt{\rho}$ and $\bar{A} = A / \sqrt{\left(\mathrm{tr}A^\ast A\right)}$.
$\bar{\psi}_{\rm F}$ and $\bar{A}_{\rm F}$ are invariant under the transformation $\mathcal{G}^{\rm n,b}(\theta, \hat{\Vec{z}}, \theta / 2)$, giving the symmetry $H_{\rm n,b} \simeq U(1) \times \mathbb{Z}_2$ for $\Vec{B} = 0$ or $\Vec{B} \parallel \hat{\Vec{z}}$.
Here, $\theta$ is the arbitrary real value.
For spinor condensates, the F state can be realized for $c_2 > 4 c_1$, $c_1 < 0$, and $q = 0$ or $c_2 > 4 c_1$, $c_1 < |q| / (2 \rho |\Vec{B}|^2)$, and $q < 0$. 
The F state appears in a certain region of the phase diagram 
in $^3P_2$ superfluids without the quasi-classical approximation 
\cite{Mizushima:2021qrz}.

In terms of invariants,
the F state can also be characterized by $\Vec{S}^2 = 4 \rho^2$ and $|\Psi_{20}|^2 = |\Psi_{30}|^2 = |\Phi_{30}|^2 = \Gamma_4 = 0$, 
shown as point F in Fig. \ref{fig:phase_image}.

%%%%%%%%%
\subsection{Canted ferromagnetic states}

The canted ferromagnetic (CF) state $\psi_{\rm CF}$ can be written as
\alignedequation{
\bar{\psi}_{\rm CF} &= \begin{pmatrix} 0 & 1 & 0 & 0 & 0 \end{pmatrix}^T, \\
\bar{A}_{\rm CF} &= - \frac{\sqrt{3}}{2} \begin{pmatrix} 0 & 0 & 1 \\ 0 & 0 & i \\ 1 & i & 0 \end{pmatrix}.
}
$\bar{\psi}_{\rm CF}$ and $\bar{A}_{\rm CF}$ are invariant under the transformation $\mathcal{G}^{\rm n,b}(\theta, \hat{\Vec{z}}, \theta)$, giving the symmetry $H_{\rm n,b} \simeq U(1)$ for $\Vec{B} = 0$ or $\Vec{B} \parallel \hat{\Vec{z}}$.

In terms of invariants,
the CF state can also be characterized by $\Vec{S}^2 = \rho^2$ and $|\Psi_{20}|^2 = |\Psi_{30}|^2 = |\Phi_{30}|^2 = \Gamma_4 = 0$, 
which is shown as point CF in Fig. \ref{fig:phase_image}.

\subsection{Nematic states}

The nematic (N) states $\psi_{\rm N}$ can be written as
\alignedequation{
\bar{\psi}_{\rm N} &= \frac{1}{\sqrt{2}} \begin{pmatrix} \cos\zeta & 0 & \sqrt{2} \sin\zeta & 0 & \cos\zeta \end{pmatrix}^T, \\
\bar{A}_{\rm N} &= \sqrt{2} \begin{pmatrix} \cos\lambda_6^+(\zeta) & 0 & 0 \\ 0 & -\cos\lambda_6^-(\zeta) & 0 \\ 0 & 0 & \sin\zeta \end{pmatrix}, \\
\lambda_n^\pm(x) &\equiv x \pm \frac{\pi}{n},
\label{eq:nematic}
}
Here $\zeta$ is the arbitrary real value.
Depending on unbroken symmetries, the nematic states can be further classified into the
$D_2$ biaxial nematic ($D_2$BN), 
$D_4$ biaxial nematic ($D_4$BN), and
uniaxial nematic (UN) states, as explained below.
The nematic states are ground-state 
$^3P_2$ superfluids 
in the weak coupling limit, 
within the quasiclassical approximation.
The nematic states are 
shown as the blue $S^1$ curve in Fig. \ref{fig:phase_image}.
The extended OPM space $(S^1 \times S^4)/\mathbb{Z}_2$ is fibered over this blue $S^1$ curve
with the fiber $U(1)\times SO(3) / D_2$ \cite{Uchino:2010}.
The fiber shrinks at the two points corresponding to the UN and $D_4$BN states.

\subsubsection{$D_2$ biaxial nematic states}

For generic $\zeta$, the nematic state is in the $D_2$BN states, $\bar{\psi}_\text{$D_2$BN}$ and $\bar{A}_\text{$D_2$BN}$, which are invariant under the transformations $\mathcal{G}^{\rm n,b}(0,\hat{\Vec{x}},\pi)$ and $\mathcal{G}^{\rm n,b}(0,\hat{\Vec{z}}, \pi)$, giving the symmetry $H_{\rm n,b} \simeq D_2$ and $H_{\rm n,b} \simeq \mathbb{Z}_2$ for $\Vec{B} = 0$ and $\Vec{B} \parallel \hat{\Vec{z}}$ respectively.
For spinor condensates, the $D_2$BN state can be realized for $c_2 < 0$, $c_2 < 4 c_1$, and $q = 0$.

In terms of invariants,
the $D_2$BN states can also be characterized by $\Vec{S}^2 = 0$, $|\Psi_{20}|^2 = \rho^2$ and $0 \leq |\Psi_{30}|^2 / \rho^3 = |\Phi_{30}|^2 / \rho^3 = \Gamma_4 / \rho^4 \leq 1$, 
which are shown as the blue curve 
(except for the two points labeled by $D_4$BN and UN) 
in Fig. \ref{fig:phase_image}.

\subsubsection{Uniaxial nematic states}

One specific case is the uniaxial nematic (UN) state.
%The UN states can be further classified into the vertical UN (VUN) state and horizontal UN (HUN) state, corresponding to specific value of $\zeta$ in Eq.~\eqref{eq:nematic}.
%
The UN state $\psi_{\rm UN}$ is the specific state of the nematic state with $\zeta = \pi/2$ in Eq.~\eqref{eq:nematic} as
\alignedequation{
\bar{\psi}_{\rm UN} &= \begin{pmatrix} 0 & 0 & 1 & 0 & 0 \end{pmatrix}^T, \\
\bar{A}_{\rm UN} &= \frac{1}{\sqrt{2}} \begin{pmatrix} -1 & 0 & 0 \\ 0 & -1 & 0 \\ 0 & 0 & 2 \end{pmatrix}.
}
$\bar{\psi}_{\rm UN}$ and $\bar{A}_{\rm UN}$ are invariant under the transformations $\mathcal{G}^{\rm n,b}(0,\hat{\Vec{x}},\pi)$ and $\mathcal{G}^{\rm n,b}(0,\hat{\Vec{z}},\theta)$, giving the symmetry $H_{\rm n,b} \simeq D_\infty \simeq O(2)$, $H_{\rm n,b} \simeq U(1)$, and $H_{\rm n,b} \simeq \mathbb{Z}_2$ for $\Vec{B} = 0$, $\Vec{B} \parallel \hat{\Vec{z}}$, and $\Vec{B} \parallel \hat{\Vec{x}}$ respectively.
For spinor condensates, the VUN state can be realized for $c_2 \lesssim 4 c_1$, $c_2 < 2 q / (\rho |\Vec{B}|^2)$, and $q > 0$.

In terms of invariants,
the UN state can also be characterized by $\Vec{S}^2 = 0$, $|\Psi_{20}|^2 / \rho^2 = |\Psi_{30}|^2 / \rho^3 = |\Phi_{30}|^2 / \rho^3 = \Gamma_4 / \rho^4 = 1$, 
which is shown as point UN in Fig. \ref{fig:phase_image}.

\subsubsection{$D_4$ biaxial nematic states}

The other specific case is the $D_4$ biaxial nematic ($D_4$BN) state, 
which is also called the square nematic state.
%We further classify the $D_4$BN states into the vertical $D_4$BN (V$D_4$BN), first horizontal $D_4$BN (H1$D_4$BN), and second horizontal $D_4$BN (H2$D_4$BN) states, corresponding to specific values of $\zeta$ in Eq.~\eqref{eq:nematic}. 
The $D_4$BN state $\psi_\text{$D_4$BN}$ is the specific state of the nematic states with $\zeta = 0$ in Eq.~\eqref{eq:nematic} as
\alignedequation{
\bar{\psi}_\text{$D_4$BN} &= \frac{1}{\sqrt{2}} \begin{pmatrix} 1 & 0 & 0 & 0 & 1 \end{pmatrix}^T, \\
\bar{A}_\text{$D_4$BN} &= \sqrt{\frac{3}{2}} \begin{pmatrix} 1 & 0 & 0 \\ 0 & -1 & 0 \\ 0 & 0 & 0 \end{pmatrix}.
}
$\bar{\psi}_\text{$D_4$BN}$ and $\bar{A}_\text{$D_4$BN}$ are invariant under the transformations $\mathcal{G}^{\rm n,b}(0,\hat{\Vec{x}},\pi)$, $\mathcal{G}^{\rm n,b}(\pi,\hat{\Vec{z}},\pi/2)$, and $\mathcal{G}^{\rm n,b}(\pi,(\hat{\Vec{x}} + \hat{\Vec{y}})/\sqrt{2},\pi)$, giving the symmetry $H_{\rm n,b} \simeq D_4$, $H_{\rm n,b} \simeq \mathbb{Z}_4$, $H_{\rm n,b} \simeq \mathbb{Z}_2$, and  $H_{\rm n,b} \simeq \mathbb{Z}_2$ for $\Vec{B} = 0$, $\Vec{B} \parallel \hat{\Vec{z}}$, $\Vec{B} \parallel \hat{\Vec{x}}$, and $\Vec{B} \parallel (\hat{\Vec{x}} + \hat{\Vec{y}})$ respectively.
For spinor condensates, the $D_4$BN state can be realized for $c_2 < 4 c_1$, $c_2 < 2 |q| / (\rho |\Vec{B}|^2)$, and $q < 0$.

In terms of invariants,
the $D_4$BN state can also be characterized by $|\Psi_{20}|^2 = \rho^2$ and $\Vec{S}^2 = |\Psi_{30}|^2 = |\Phi_{30}|^2 = \Gamma_4 =0$, 
which is shown as point $D_4$BN  in Fig. \ref{fig:phase_image}.

\subsection{Cyclic state}

%Cyclic (C) states can be classified into 
%the vertical cyclic (VC) and 
%horizontal cyclic (HC) states,
%depending on 
%the eigenvalues of
%the order parameters
%relative to the direction of the magnetic field.

The cyclic (C) state $\psi_{\rm C}$ is written as
\alignedequation{
\bar{\psi}_{\rm C} &= \frac{1}{\sqrt{3}} \begin{pmatrix} 1 & 0 & 0 & \sqrt{2} & 0 \end{pmatrix}^T, \\
\bar{A}_{\rm C} &= \frac{1}{2} \begin{pmatrix} 1 & i & \sqrt{2} \\ i & -1 & -\sqrt{2}i \\ \sqrt{2} & -\sqrt{2}i & 0 \end{pmatrix}.
}
$\bar{\psi}_{\rm C}$ and $\bar{A}_{\rm C}$ are invariant under the transformations $\mathcal{G}^{\rm n,b}(0,(\sqrt{2} \hat{\Vec{x}} + \hat{\Vec{z}})/ \sqrt{3},\pi)$ and $\mathcal{G}^{\rm n,b}(- 2 \pi / 3,\hat{\Vec{z}},2 \pi/3)$, giving the symmetry $H_{\rm n,b} \simeq T$, $H_{\rm n,b} \simeq \mathbb{Z}_3$, and $H_{\rm n,b} \simeq \mathbb{Z}_2$ for $\Vec{B} = 0$, $\Vec{B} \parallel \hat{\Vec{z}}$, and $\Vec{B} \parallel (\sqrt{2} \hat{\Vec{x}} + \hat{\Vec{z}})$ respectively.

The other well-known form of the C state is
\alignedequation{
\mathcal{G}_{\rm C}^{\rm b} \bar{\psi}_{\rm C} &= \frac{1}{2} \begin{pmatrix} i & 0 & \sqrt{2} & 0 & i \end{pmatrix}^T, \\
\mathcal{G}_{\rm C}^{\rm n} \bar{A}_{\rm C} &= \begin{pmatrix} e^{2 i \pi / 3} & 0 & 0 \\ 0 & e^{- 2 i \pi / 3} & 0 \\ 0 & 0 & 1 \end{pmatrix},
}
where $\mathcal{G}_{\rm C}^{\rm n,b} \equiv \mathcal{G}^{\rm n,b}(0,\hat{\Vec{z}},-\pi/4) \mathcal{G}^{\rm n,b}(0,\hat{\Vec{y}},-\cos^{-1}(1/\sqrt{3}))$.
In this form, $\mathcal{G}_{\rm C}^{\rm b} \bar{\psi}_{\rm HC}$ and $\mathcal{G}_{\rm C}^{\rm n} \bar{A}_{\rm HC}$ are invariant under the transformations $\mathcal{G}^{\rm n,b}(- 2 \pi / 3,(\hat{\Vec{x}} + \hat{\Vec{y}} + \hat{\Vec{z}})/ \sqrt{3},2\pi/3)$ and $\mathcal{G}^{\rm n,b}(0,\hat{\Vec{z}},\pi)$, giving the symmetry $H_{\rm n,b} \simeq \mathbb{Z}_2$ and $H_{\rm n,b} \simeq \mathbb{Z}_3$ for $\Vec{B} \parallel \hat{\Vec{z}}$ and $\Vec{B} \parallel (\hat{\Vec{x}} + \hat{\Vec{y}}+ \hat{\Vec{z}})$ respectively.
For spinor condensates, the cyclic state can be realized for $c_1 > 0$, $c_2 > 0$, and $q = 0$.

In terms of invariants,
the C state can also be characterized by $|\Psi_{30}|^2 = 2 \rho^3$ and $\Vec{S}^2 = |\Psi_{20}|^2 = |\Phi_{30}|^2 = \Gamma_4 = 0$,  
which is shown as point C in Fig. \ref{fig:phase_image}.

\subsection{Canted cyclic states}

The canted cyclic (CC) states $\psi_{\rm CC}$ are intermediate states interpolating C, UN, and $D_4$BN states written as
\alignedequation{
\bar{\psi}_{\rm CC} &= \frac{1}{\sqrt{2}} \begin{pmatrix} i \cos\eta & 0 & \sqrt{2} \sin\eta & 0 & i \cos\eta \end{pmatrix}^T, \\
\bar{A}_{\rm CC} &= \sqrt{2} \begin{pmatrix} \sigma_\eta & 0 & 0 \\ 0 & \sigma_\eta^\ast & 0 \\ 0 & 0 & \sin\eta \end{pmatrix}, \\
\sigma_\eta &= \sin\eta \cos(2 \pi/3) + i \cos\eta \sin(2 \pi/3).
}
The 
C, UN, and $D_4$BN states are the specific states with $\eta = \pi/4$, $\pi/2$, and $0$, respectively.
$\bar{\psi}_{\rm CC}$ and $\bar{A}_{\rm CC}$ are invariant under the transformations $\mathcal{G}^{\rm n,b}(0,\hat{\Vec{x}},\pi)$ and $\mathcal{G}^{\rm n,b}(0,\hat{\Vec{z}},\pi)$, giving the symmetry $H_{\rm n,b} \simeq D_2$ and $H_{\rm n,b} \simeq \mathbb{Z}_2$ for $\Vec{B} = 0$ and $\Vec{B} \parallel \hat{\Vec{z}}$.
For spinor condensates, the CC state can be realized for $c_2 < 4 c_1$, $c_1 > |q| / (2 \rho |\Vec{B}|^2)$, $c_2 > 2 |q| / (\rho |\Vec{B}|^2)$, and $q \neq 0$.

In terms of the invariants,
the CC state can also be characterized by $\Vec{S}^2 = 0$, $0 \leq |\Psi_{20}|^2 \leq \rho^2$, and
\alignedequation{
|\Psi_{30}|^2 &= 2 \rho^3- \frac{3 |\Psi_{20}|^4}{2 \rho} - \frac{|\Psi_{20}|^6}{2 \rho^3}, \\
|\Phi_{30}|^2 &= \frac{|\Psi_{20}|^4}{2 \rho} - \frac{|\Psi_{20}|^6}{2 \rho^3}, \\
\Gamma_4 &= - \frac{|\Psi_{20}|^6}{2 \rho^3} + \frac{|\Psi_{20}|^8}{2 \rho^5},
}
which is shown as the green circle in Fig. \ref{fig:phase_image}.

%\begin{color}{red}
\subsection{Reduced ferromagnetic states}

The
reduced ferromagnetic (RF) states $\psi_{\rm RF}$ are the intermediate states interpolating F and $D_4$BN states:
\alignedequation{
\bar{\psi}_{\rm RF} &= \begin{pmatrix} \cos\nu & 0 & 0 & 0 & \sin\nu \end{pmatrix}^T, \\
\bar{A}_{\rm RF} &= \frac{\sqrt{6}}{2} \begin{pmatrix} \sin\lambda_4^+(\nu) & i \sin\lambda_4^-(\nu) & 0 \\ i \sin\lambda_4^-(\nu) & - \sin\lambda_4^+(\nu) & 0 \\ 0 & 0 & 0 \end{pmatrix}.
}
The F and $D_4$BN states are the specific states with $\nu = 0$ and $\pi/4$, respectively.
$\bar{\psi}_{\rm RF}$ and $\bar{A}_{\rm RF}$ are invariant under the transformation $\mathcal{G}^{\rm n,b}(\pi,\hat{\Vec{z}},\pi/2)$, giving the symmetry $H_{\rm n,b} \simeq \mathbb{Z}_4$ for $\Vec{B} = 0$ or $\Vec{B} \parallel \hat{\Vec{z}}$.

In terms of the invariants,
the RF states can also be characterized by $0 \leq \Vec{S}^2 = 4 (\rho^2 - |\Psi_{20}|^2) \leq 4 \rho^2$, $|\Psi_{30}|^2 = |\Phi_{30}|^2 = \Gamma_4 = 0$, which are shown as the green 
curve in Fig.~\ref{fig:phase_image}.

\subsection{Reduced canted ferromagnetic state}

The reduced canted ferromagnetic (RCF) states $\psi_{\rm RCF}$ are intermediate states interpolating the CF and $D_4$BN states:
\alignedequation{
\!\!\!\!\!\!\bar{\psi}_{\rm RCF} &= \begin{pmatrix} 0 & \cos\xi & 0 & \sin\xi & 0 \end{pmatrix}^T, \\
\!\!\!\!\!\!\bar{A}_{\rm RCF} &= - \frac{\sqrt{6}}{2} \begin{pmatrix} 0 & 0 & \sin\lambda_4^-(\xi) \\ 0 & 0 & i \sin\lambda_4^+(\xi) \\ \sin\lambda_4^-(\xi) & i \sin\lambda_4^+(\xi) & 0 \end{pmatrix}.
}
The F and $D_4$BN states are the specific states with $\xi = 0$ and $\pi/4$, respectively.
$\bar{\psi}_{\rm RCF}$ and $\bar{A}_{\rm RCF}$ are invariant under the transformation $\mathcal{G}^{\rm n,b}(\pi,\hat{\Vec{z}},\pi)$, giving the symmetry $H_{\rm n,b} \simeq \mathbb{Z}_2$ for $\Vec{B} = 0$ or $\Vec{B} \parallel \hat{\Vec{z}}$.

In terms of the invariants,
the RF states can also be characterized by $0 \leq \Vec{S}^2 = \rho^2 - |\Psi_{20}|^2 \leq \rho^2$, $|\Psi_{30}|^2 = |\Phi_{30}|^2 = \Gamma_4 = 0$, shown as the green curve in Fig. \ref{fig:phase_image}.

\subsection{Mixed state}

The mixed (M) states $\psi_{\rm M}$ are intermediate states interpolating the C, F, and CF states, which can be written as
\alignedequation{
\bar{\psi}_{\rm M} &= \begin{pmatrix} \cos\phi & 0 & 0 & \sin\phi & 0 \end{pmatrix}^T, \\
\bar{A}_{\rm M} &= \frac{\sqrt{3}}{2} \begin{pmatrix} \cos\phi & i \cos\phi & \sin\phi \\ i \cos\phi & - \cos\phi & - i \sin\phi \\ \sin\phi & -i\sin\phi & 0 \end{pmatrix}.
}
The C, F, and CF states are the specific states with $\phi = \cos^{-1}(1/\sqrt{3})$, $\phi=0$, and $\phi=\pi/2$, respectively.
$\bar{\psi}_{\rm M}$ and $\bar{A}_{\rm M}$ are invariant under the transformation $\mathcal{G}^{\rm n,b}(- 2 \pi/ 3,\hat{\Vec{z}},2 \pi/3)$, giving the symmetry $H_{\rm n,b} \simeq \mathbb{Z}_3$ for $\Vec{B} = 0$ or $\Vec{B} \parallel \hat{\Vec{z}}$.
For spinor condensates, the M state can be realized for $c_2 > 4 c_1$, $c_1 > |q| / (2 \rho |\Vec{B}|^2)$, $c_2 > 2 |q| / (\rho |\Vec{B}|^2)$, and $q < 0$.

In terms of the invariants,
the M states can also be characterized by
$- \rho^2 \leq (\Vec{S}^2)^{1/2} \leq 4 \rho^2$, $|\Psi_{20}|^2 = |\Phi_{30}|^2 = 0$, and
\begin{align}
|\Psi_{30}|^2 = \frac{4 \rho^3 - 3 \rho \Vec{S}^2 + (\Vec{S}^2)^{3/2}}{2},
\end{align}
which are shown as the green curve in Fig.~\ref{fig:phase_image}.

\subsection{Broken axisymmetric states}

%Broken axisymmetric (BA) states can be classified into 
%the vertical broken axisymmetric (VBA) and 
%horizontal broken axisymmetric (HBA) states,
%depending on 
%the eigenvalues of
%the order parameters
%relative to the direction of the magnetic field.

The broken axisymmetric (BA) states are intermediate states  interpolating the RF and CC states, which can be written as
\alignedequation{
\bar{\psi}_{\rm BA} &= \begin{pmatrix} a+b & 0& c & 0 & a-b \end{pmatrix}^T, \\
\bar{A}_{\rm BA} &= \frac{1}{\sqrt{2}} \begin{pmatrix} \sqrt{6} a - c & \sqrt{6} i b & 0 \\ \sqrt{6} i b & - \sqrt{6} a - c & 0 \\ 0 & 0 & 2 c \end{pmatrix}. \\
}
Here, $a$, $b$, and $c$ are arbitrary real values satisfying $1 = 2 (a^2 + b^2) + c^2$.
RF and CC states are the specific states with $c = 0$ and $a = 0$, respectively.
$\bar{\psi}_{\rm BA}$ and $\bar{A}_{\rm BA}$ are invariant under the transformation or $\mathcal{G}^{\rm n,b}(0,\hat{\Vec{z}},\pi)$, giving the symmetry $H_{\rm n,b} \simeq \mathbb{Z}_2$ for $\Vec{B} = 0$ or $\Vec{B} \parallel \hat{\Vec{z}}$.
For spinor condensates, the BA state can be realized for $c_2 \gtrsim 4 c_1$, $q > 0$ and $\Vec{B} \parallel \hat{\Vec{x}}$ where the symmetry becomes trivial $H_{\rm n,b} \simeq 1$.
The BA states (also called magnetized BN states in Ref.~\cite{Mizushima:2021qrz}) appear in a certain region of the phase diagram
for $\Vec{B} \parallel \hat{\Vec{z}}$ in $^3P_2$ superfluids without the quasi-classical approximation 
\cite{Mizushima:2021qrz}.

In terms of the invariants,
the BA states can also be characterized by $0 \leq \Vec{S}^2 \leq 4 \rho^2$, $- (\rho^2 - \Vec{S}^2 / 4) \leq (|\Psi_{20}|^2)^{1/2} \leq \rho^2 - \Vec{S}^2 / 4$, and
\alignedequation{
\!\!\!\!\!\!|\Psi_{30}|^2 &= \frac{\chi \rho^2 \{4 \rho^2 - \Vec{S}^2 + 6 (|\Psi_{20}|^2)^{1/2} \rho + 2 |\Psi_{20}|^2\}^2}{32 \{\rho + (|\Psi_{20}|^2)^{1/2}\}^3}, \\
\!\!\!\!\!\!|\Phi_{30}|^2 &= \frac{\chi \rho^2 \{\Vec{S}^2 + 2 (|\Psi_{20}|^2)^{1/2} \rho + 2 |\Psi_{20}|^2\}^2}{32 \{\rho + (|\Psi_{20}|^2)^{1/2}\}^3}, \\
\!\!\!\!\!\!\Gamma_4 &= (|\Psi_{20}|^2)^{1/2} |\Phi_{30}|^2, \\
\!\!\!\!\!\!\chi &\equiv 4 - \frac{\Vec{S}^2}{\rho^2} - \frac{4 |\Psi_{20}|^2}{\rho^2},
\label{eq:BA-constraint}
}
which are shown as the gray surface in Fig.~\ref{fig:phase_image}.

\subsection{Phase diagram and topological structure for symmetric uniform state}

%\begin{color}{red}
%Let us discuss again $\cdots$
\begin{figure}[htb]
\centering
\begin{minipage}{0.49\linewidth}
\centering
\includegraphics[width=0.99\linewidth]{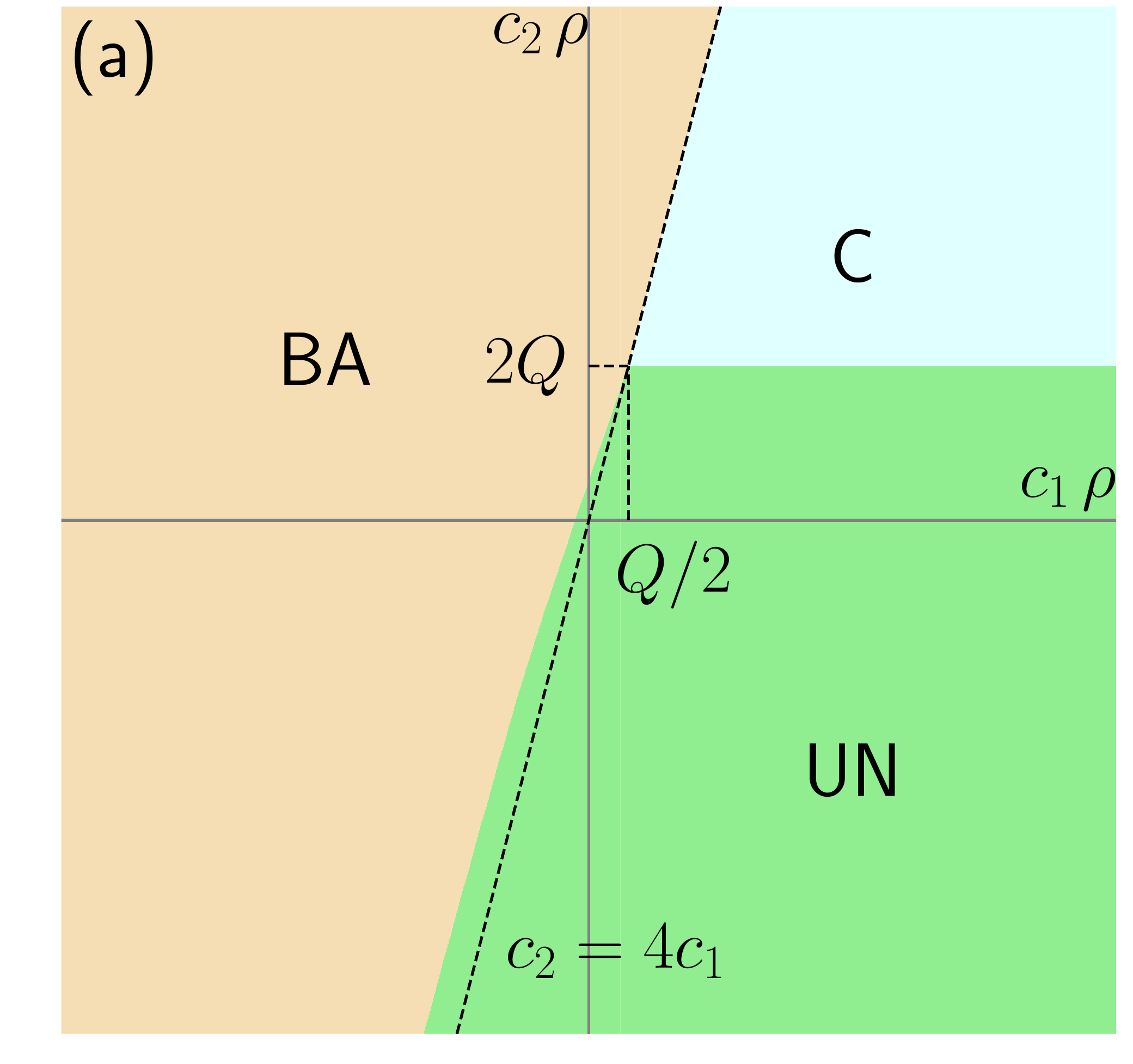}
\end{minipage}
\begin{minipage}{0.49\linewidth}
\centering
\includegraphics[width=0.99\linewidth]{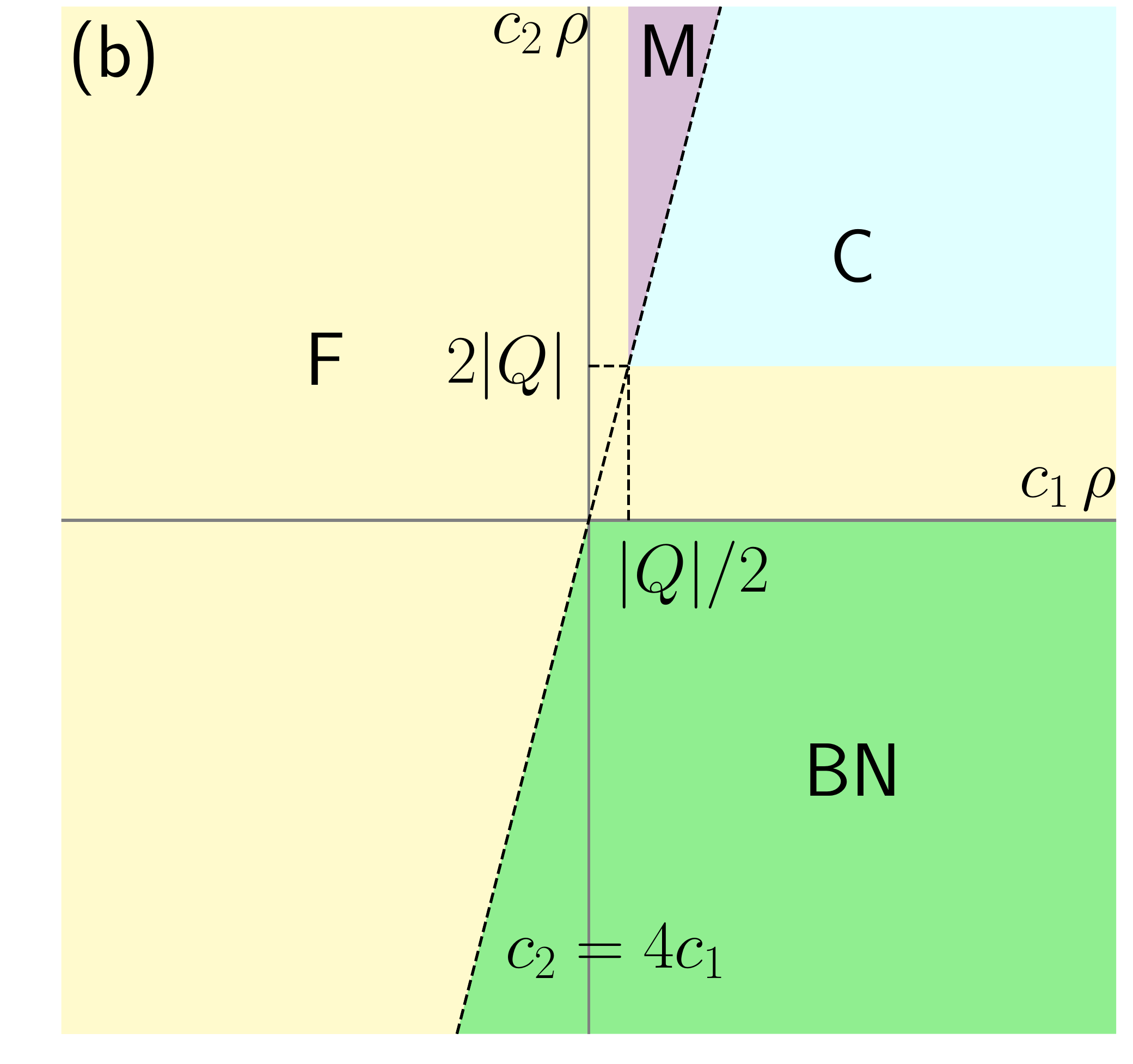}
\end{minipage} \\
\begin{minipage}{0.49\linewidth}
\centering
\includegraphics[width=0.99\linewidth]{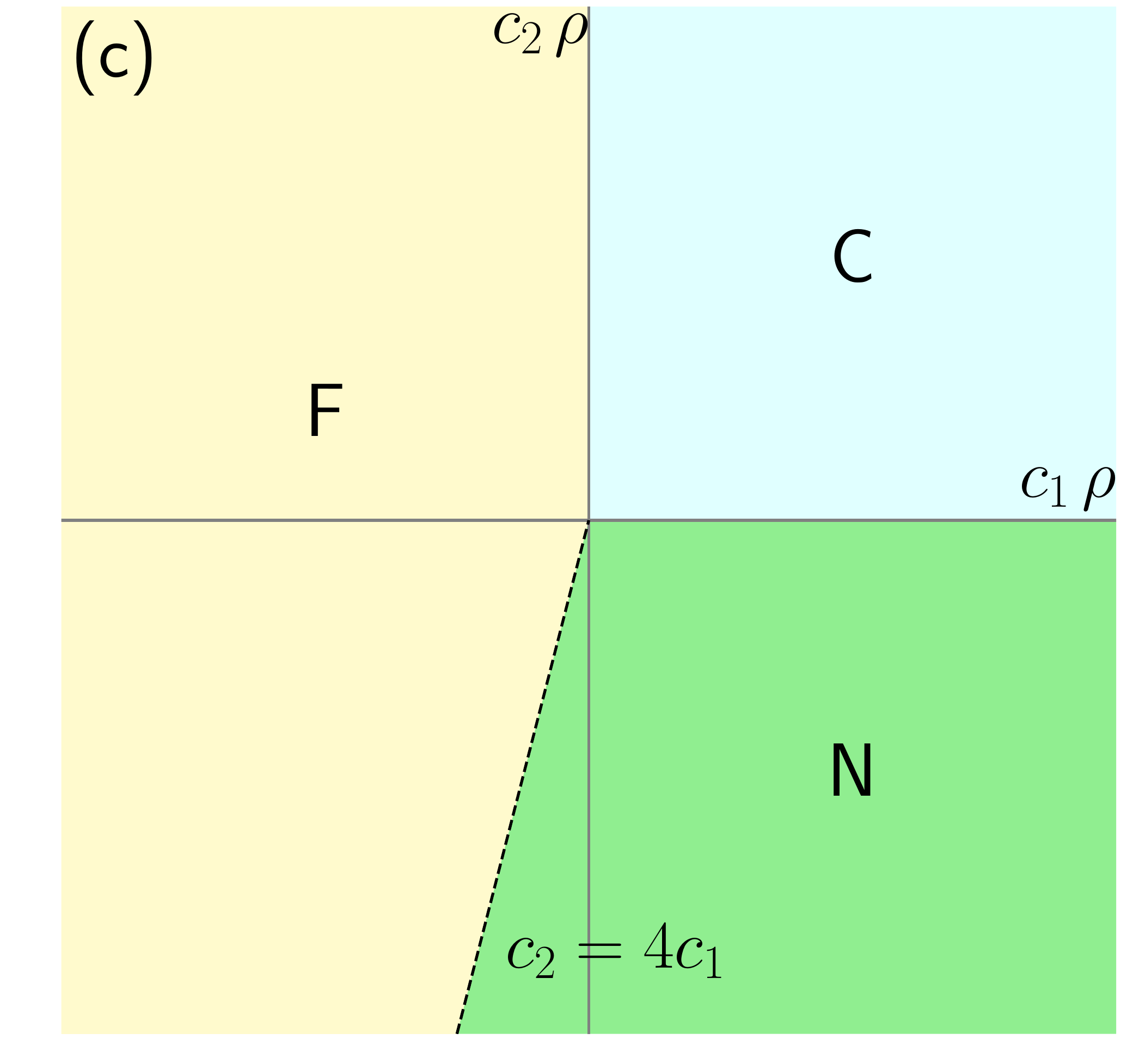}
\end{minipage}
\begin{minipage}{0.49\linewidth}
\centering
\includegraphics[width=0.99\linewidth]{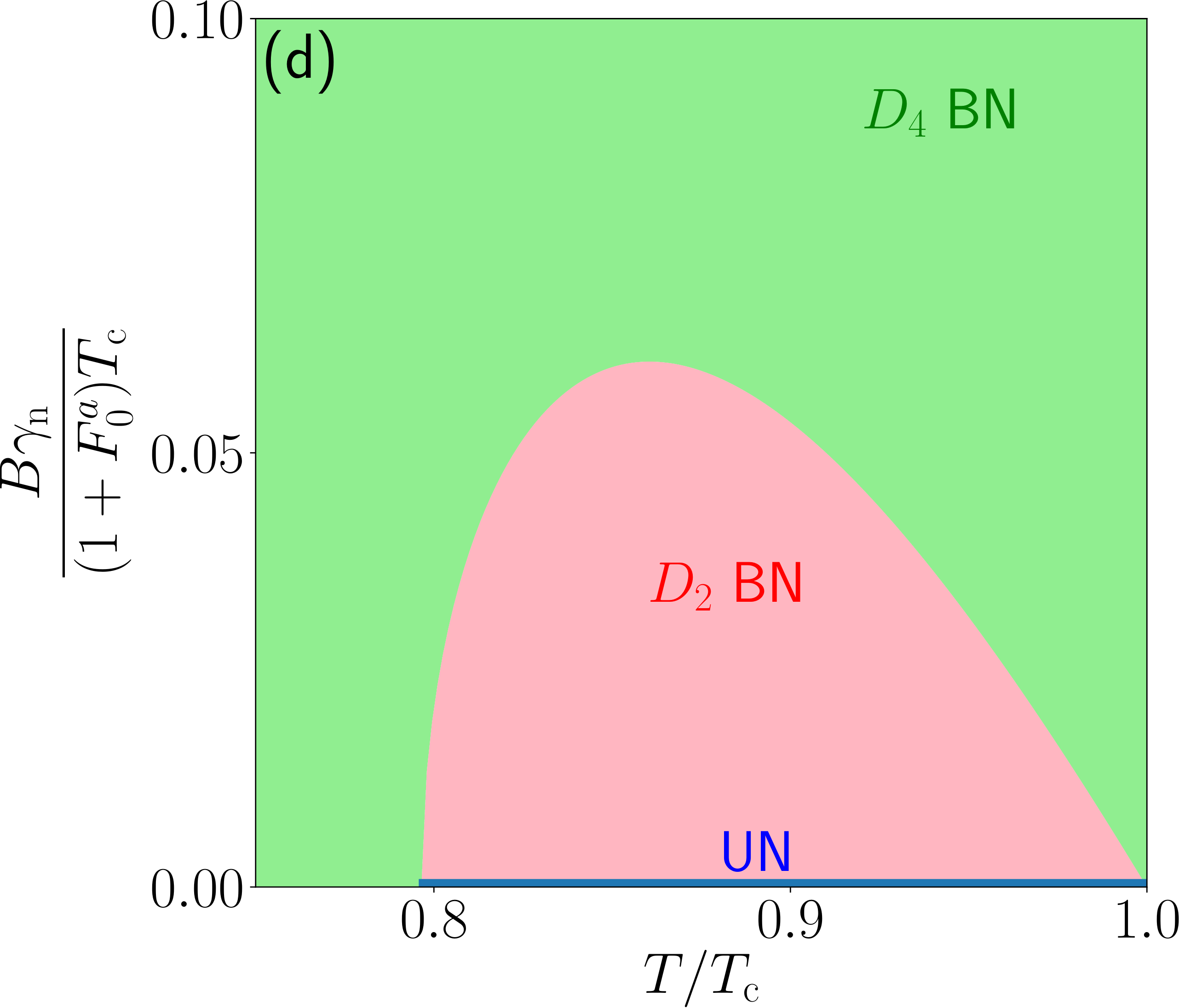}
\end{minipage} \\
\caption{
\label{fig:phase_diagram}
Phase diagram for the spinor condensates (a--c) and the neutron $^3P_2$ superfluid (d).
In panels (a), (b), and (c), the coefficient of the quadratic Zeeman energy $Q \equiv q |\Vec{B}|^2$ is positive, negative, and zero, respectively.
In panel (a), the boundary between the BA and UN states can be numerically determined.
In panel (d), the UN ($B = 0$) and $D_2$BN ($B > 0$) states appear only at $T \gtrsim 0.796$.
The critical magnetic field between $D_2$BN and $D_4$BN states takes the maximal $B = 6.05 \times 10^{-3} (1 + F_0^a) T_{\rm c} / \gamma_{\rm n}$ at $T = 0.860 T_{\rm c}$.
}
\end{figure}
In Fig. \ref{fig:phase_diagram}, we show phase diagrams for the spinor condensates for the Hamiltonian density $h$ in Eq. \eqref{eq:spinor-Hamiltonian-original}
and neutron superfluid for the free energy density $f$ in Eq. \eqref{eq:neutron_Free_energy}.
For spinor condensates, the F, M, C, and BN (BA, C, and UN) states appear for $q > 0$ ($q < 0$).
In the case of $q = 0$, the F, C, and N states appear, and 
all states ($D_2$BN, $D_4$BN, and UN states) in the N states 
are degenerated.

For neutron $^3P_2$ superfluids, the UN, $D_2$BN, and $D_4$BN states are dominant within the present framework, which corresponds to the region with positive $c_1$ and negative $c_2$ for the spinor condensates.
At high temperatures, the $D_4$BN and $D_2$BN (UN) states appear for $\Vec{B} \neq 0$ ($\Vec{B} = 0$).
At low temperatures, the only $D_4$BN state appears, which contradicts the result obtained by the BdG analysis \cite{Mizushima:2016fbn}
giving the $D_4$BN, $D_2$BN, and UN states at any temperature.
This contradiction originates from the fact that 
the GL expansion is valid only around the critical temperature, 
and this discrepancy may be cured by including 
 higher order terms into
the free-energy density $f$ in 
Eq.~\eqref{eq:neutron_Free_energy}
which neglects terms higher than the 8th order.
Besides these three states, the existences of the F and BA states 
\cite{2010arXiv1001.2072K} 
are also predicted 
\cite{Mizushima:2021qrz}
without the quasi-classical approximation that we use to obtain the free energy density $f$ in Eq. \eqref{eq:neutron_Free_energy} from the Lagrangian $\mathcal{L}$ in Eq. \eqref{eq:neutron_Lagrangian}

Now let us go back to 
Fig. \ref{fig:phase_image} in which we have shown 
all the states having the nontrivial symmetry $H_{\rm n,b}$ under $\Vec{B} = 0$ explained above.
The BA states for $2a^2 + 2b^2 + c^2 = 1$ are represented by 
the gray $S^2$ surface, and the $D_2$BN, CC, RF states are denoted by the green, blue and purple $S^1$ curves parametrized by $\zeta$, $\eta$, and $\nu$, respectively.
%, on the $S^2$ surface for the BN state.
The $S^2$ surface can also be constructed with the $U(1) \times SO(3)$-invariants
$\{\Vec{S}^2,  |\Psi_{20}|^2, |\Psi_{30}|^2, |\Phi_{30}|^2, \Gamma_4\}$ under the constraints shown in Eq.~\eqref{eq:BA-constraint}.
The RCF state is represented by the $S^1$ curves for $\xi$ out of the $S^2$ surface for the BN state except for the point corresponding to $D_4$BN states.
The M state is represented by the $S^1$ curves for $\phi$ out of the $S^2$ surface for the BN state except for two points corresponding to C and F states.
The F, UN, and $D_4$BN states correspond to the points on the $S^2$ surface for the BN state, while the CF state corresponds to the point out of the $S^2$ surface for the BN state. 
%\end{color}

%%%%%%%%%%%%%%
\section{$^3P_2$ neutron superfluids}\label{sec:3P2}

\begin{table*}[htb]
\centering
\begin{tabular}{ccccccc} \hline\hline
State & $\Vec{S}^2/\rho^2$ & $|\Psi_{20}|^2/\rho^2$ & $|\Psi_{30}|^2/\rho^3$ & $|\Phi_{30}|^2 / \rho^3$ & $\Gamma_4 / \rho^4$ & $H_{\rm n,b}$ \\ \hline
F & 4 & 0 & 0 & 0 & 0 & $[U(1)\times\mathbb{Z}_2]_{\Vec{B}=0,\, \Vec{B}\parallel\hat{\Vec{z}}}$ \\
CF & 1 & 0 & 0 & 0 & 0 & $[U(1)]_{\Vec{B}=0,\, \Vec{B}\parallel\hat{\Vec{z}}}$ \\
$D_2$BN  & 0 & 1 & $\sin^2(3\zeta)$ & $\sin^2(3\zeta)$ & $\sin^2(3 \zeta)$ & $[D_2]_{\Vec{B}=0}$, $[\mathbb{Z}_2]_{\Vec{B}\parallel\hat{\Vec{z}}}$ \\
UN & 0 & 1 & 1 & 1 & 1 & $[D_\infty]_{\Vec{B}=0}$, $[U(1)]_{\Vec{B}\parallel\hat{\Vec{z}}}$, $[\mathbb{Z}_2]_{\Vec{B}\parallel\hat{\Vec{x}}}$ \\
$D_4$BN & 0 & 1 & 0 & 0 & 0 & $[D_4]_{\Vec{B}=0}$, $[\mathbb{Z}_4]_{\Vec{B}\parallel\hat{\Vec{z}}}$, $[\mathbb{Z}_2]_{\Vec{B}\parallel\hat{\Vec{x}},\,\Vec{B}\parallel(\hat{\Vec{x}} + \hat{\Vec{y}})}$ \\
C & 0 & 0 & 2 & 0 & 0 & $[T]_{\Vec{B}=0}$, $[\mathbb{Z}_3]_{\Vec{B}\parallel\hat{\Vec{z}}}$, $[\mathbb{Z}_2]_{\Vec{B}\parallel(\sqrt{2}\hat{\Vec{x}}+\hat{\Vec{z}})}$ \\
CC & 0 & $\cos^2(2\eta)$ & $ \frac{1}{2} F^{(3)}_{4,0,-3,-1}(\eta)$ & $ \frac{1}{2} F^{(3)}_{0,0,1,-1}(\eta)$ & $ \frac{1}{2} F^{(4)}_{0,0,0,-1,1}(\eta)$ & $[D_2]_{\Vec{B}=0}$, $[\mathbb{Z}_2]_{\Vec{B}\parallel\hat{\Vec{z}}}$ \\
RF & $4 \cos^2(2\nu)$ & $\sin^2(2\nu)$ & 0 & 0 & 0 & $[\mathbb{Z}_4]_{\Vec{B}=0,\,\Vec{B}\parallel\hat{\Vec{z}}}$ \\
RCF & $\cos^2(2\xi)$ & $\sin^2(2\xi)$ & 0 & 0 & 0 & $[\mathbb{Z}_2]_{\Vec{B}=0,\,\Vec{B}\parallel\hat{\Vec{z}}}$ \\
M & $ \frac{1}{4} F^{(2)}_{1,6,9}(\phi)$ & 0 & $ \frac{27}{16} F^{(3)}_{1,-1,-1,1}(\phi)$ & 0 & 0 & $[\mathbb{Z}_3]_{\Vec{B}=0,\,\Vec{B}\parallel\hat{\Vec{z}}}$ \\
BA & $64 a^2 b^2$ & $(1 - 4 b^2)^2$ & $F_{ab}^+ c^2$ & $F_{ab}^- c^2$ & $(1 - 4 b^2) F_{ab}^- c^2$ & $[\mathbb{Z}_2]_{\Vec{B}=0,\,\Vec{B}\parallel\hat{\Vec{z}}}$ \\
\end{tabular}
\caption{
\label{table:terms}
$U(1) \times SO(3)$ invariant terms $\Vec{S}^2$, $|\Psi_{20}|^2$, $|\Psi_{30}|^2$, $|\Phi_{30}|^2$, and $\Gamma_4$, and unbroken symmetry $H_{\rm n,b}$ for uniform ground states.
Here $F(x)$ defined as the expansion with $\cos(2 x)$: $F^{(n)}_{a_0, \cdots a_n}(x) \equiv \sum_{k=0}^{n} a_k \cos^k(2 x)$ and $F_{ab}^\pm \equiv (1 - 8 a^2 \pm 4 b^2)^2$.
}
\end{table*}

\begin{table*}[htb]
\centering
\begin{tabular}{cccccc} \hline\hline
State & $f_{004}/\rho^2$ & $f_{006}/\rho^3$ & $f_{008}/\rho^4$ & $f_{022}/(\rho B^2)$ & $f_{024}/(\rho^2 B^2)$ \\ \hline
$[$F$]_{\Vec{B}\parallel\hat{\Vec{z}}}$ & $9$ & \underline{$-648$} & $15552$ & \underline{$0$} & $0$ \\
$[$CF$]_{\Vec{B}\parallel\hat{\Vec{z}}}$ & $\frac{27}{4}$ & $-405$ & $8505$ & $\frac{3}{2}$ & $- 45$ \\
$[$UN$]_{\Vec{B}\parallel\hat{\Vec{z}}}$ & \underline{$\frac{9}{2}$} & $-174$ & $2376$ & $2$ & \underline{$-84$} \\
$[D_4$BN$]_{\Vec{B}\parallel\hat{\Vec{z}}}$ & \underline{$\frac{9}{2}$} & $-162$ & \underline{$1944$} & \underline{$0$} & $0$ \\
$[D_4$BN$]_{\Vec{B}\parallel\hat{\Vec{x}},\,\Vec{B}\parallel(\hat{\Vec{x}}+\hat{\Vec{y}})}$ & \underline{$\frac{9}{2}$} & $-162$ & \underline{$1944$} & $3/2$ & $-54$ \\
$[D_2$BN$]_{\Vec{B}\parallel\hat{\Vec{z}}}$ & \underline{$\frac{9}{2}$} & $-6 F^{(3)}_{28,3,0,-4}(\zeta)$ & $216 F^{(3)}_{10,3,0,-4}(\zeta)$ & $F^{(1)}_{1,-1}(\zeta)$ & $- 6 F^{(2)}_{5,-7,2}(\zeta)$ \\
$[$C$]_{\Vec{B}\parallel\hat{\Vec{z}},\,\Vec{B}\parallel(\sqrt{2}\hat{\Vec{x}}+\hat{\Vec{z}})}$ & $6$ & $-294$ & $4752$ & $1$ & $-36$ \\
$[$CC$]_{\Vec{B}\parallel\hat{\Vec{z}}}$ & $\frac{3}{2} F^{(2)}_{4,0,-1}(\eta)$ & $-6 F^{(3)}_{49,0,-21,-1}(\eta)$ & $216 F^{(3)}_{22,0,-12,-1}(\eta)$ & $F^{(1)}_{1,-1}(\eta)$ & $- 6 F^{(2)}_{6,-7,1}(\eta)$ \\
$[$RF$]_{\Vec{B}\parallel\hat{\Vec{z}}}$ & $\frac{9}{2} F^{(2)}_{1,0,1}(\nu)$ & $-162 F^{(2)}_{1,0,3}(\nu)$ & $1944 F^{(4)}_{1,0,6,0,1}(\nu)$ & $0$ & $0$ \\
$[$RCF$]_{\Vec{B}\parallel\hat{\Vec{z}}}$ & $\frac{9}{4} F^{(2)}_{2,0,1}(\xi)$ & $-81 F^{(2)}_{2,0,3}(\xi)$ & $243 F^{(4)}_{8,0,24,0,3}(\xi)$ & $3/2$ & $- 9 F^{(2)}_{5,0,-1}(\xi)$ \\
$[$M$]_{\Vec{B}\parallel\hat{\Vec{z}}}$ & $\frac{9}{16} F^{(2)}_{11,2,3}(\phi)$ & $-\frac{81}{16} F^{(3)}_{63,29,41,-5}(\phi)$ & $\frac{243}{16} F^{(4)}_{363,292,402,-60,27}(\phi)$ & $\frac{3}{4} F^{(1)}_{1,-1}(\phi)$ & $- \frac{9}{4} F^{(2)}_{13,-10,-3}(\phi)$ \\
$[$BA$]_{\Vec{B}\parallel\hat{\Vec{z}}}$ & Eq.~\eqref{eq:BA-f004} & Eq.~\eqref{eq:BA-f006} & Eq.~\eqref{eq:BA-f008} & Eq.~\eqref{eq:VBA-f022} & Eq.~\eqref{eq:VBA-f024} \\
$[$BA$]_{\Vec{B}\parallel\hat{\Vec{x}}}$ & Eq.~\eqref{eq:BA-f004} & Eq.~\eqref{eq:BA-f006} & Eq.~\eqref{eq:BA-f008} & Eq.~\eqref{eq:HBA-f022} & Eq.~\eqref{eq:HBA-f024}
\end{tabular}
\caption{
\label{table:ground-state}
Components of free-energy density $f$ for uniform ground states.
The lowest values for each term are highlighted with underlining.
}
\end{table*}

%Now, we can investigate the stability for neutron superfluidity by calculating $f$ in Eq. \eqref{eq:neutron_terms} for each states.
When $\Vec{B} = 0$, uniform part of the free-energy density $\alpha f_{002} + \beta_0 f_{004} + \gamma_0 f_{006} + \delta_0 f_{008}$ and the Hamiltonian density $h_{\rm int}$ includes only terms which are invariant under the $U(1) \times SO(3)$ transformation $\mathcal{G}_{\rm n,b}$.
Because the internal degrees of freedom for $A$ or $\psi$ are nine 
after fixing $\rho$, there are five independent $U(1) \times SO(3)$ invariant terms.
We fix them as $\Vec{S}^2$, $|\Psi_{20}|^2$, $|\Psi_{30}|^2$, $|\Phi_{30}|^2$, and $\Gamma_4$.
In Table \ref{table:terms}, we summarize them and unbroken symmetries in the free-energy density $f$ for each uniform ground state.
In Table \ref{table:ground-state}, we also summarize components $\Vec{S}^2$, $|\Psi_{20}|^2$, $|\Psi_{30}|^2$, and $|\Phi_{30}|^2$ and components in the free-energy density $f$ for each uniform ground state, respectively.
For the BA state, some results are too long to be shown in the table, and we write them here:
\begin{align}
& \begin{aligned}
\frac{f_{004}}{\rho^2} = \frac{9}{2} + 12 (1 + 4 a^2 - 2 b^2) b^2,
\label{eq:BA-f004}
\end{aligned} \\
& \begin{aligned}
\frac{f_{006}}{\rho^3} &=  - 6 \{ 29 - 4 a^2 (3 - 8 a^2)^2 + 156 b^2 \\
&\hspace{-17pt}+ 48 a^2 (31 - 24 a^2) b^2 - 96 (3 + 10 a^2) b^4 - 64 b^6 \},
\label{eq:BA-f006}
\end{aligned} \\
& \begin{aligned}
\frac{f_{008}}{\rho^4} &= 216 \{ 11 - 4 a^2 (3 - 8 a^2)^2 + 84 b^2 \\
&\hspace{-15pt}+ 48 a^2 (21 + 16 a^2 - 64 a^4) b^2 \\
&\hspace{-15pt}- 48 (3 - 28 a^2 +  96 a^4) b^4 - 64 (1 + 60 a^2) b^6\}.
\label{eq:BA-f008}
\end{aligned}
\end{align}
We further show 
\begin{align}
& \frac{f_{022}}{\rho B^2} = 2 c^2, \label{eq:VBA-f022} \\
& \frac{f_{024}}{\rho^2 B^2} = -12 (6 a^2 + 10 b^2 + 7 c^2) c^2 \label{eq:VBA-f024},
\end{align}
for $\Vec{B} \parallel \hat{\Vec{z}}$, and
\begin{align}
&\!\!\! \frac{f_{022}}{\rho B^2} = \frac{3}{2} - (\sqrt{6} a + c) c, \label{eq:HBA-f022} \\
\begin{split}
&\!\!\! \frac{f_{024}}{\rho^2 B^2} = - 6 \{ 2 + 22 a^2 + 18 b^2 - 8 (a^2 - b^2) (2 a^2 - b^2) \\
&\phantom{\frac{f_{024}}{\rho^2 B^2} = - 6} - \sqrt{6} a c (8 a^2 - 4 b^2 + 5) \},
\end{split}
\label{eq:HBA-f024}
\end{align}
for $\Vec{B} \parallel \hat{\Vec{x}}$.

In Table~\ref{table:ground-state}, we also highlight the lowest values for each terms.
At the temperature close to the superfluid transition temperature $T_{\rm c}$, the N state ($D_2$BN, UN, and $D_4$BN states) has the lowest value of $f_{004} = (9/2) \rho^2$, and the UN ($D_4$BN) state has the lower value of $f_{006} = -174 \rho^3$ ($f_{008} = 1944\rho^4$) than other nematic states.
Under the magnetic field, the $D_4$BN (UN) state with $\Vec{B} \parallel \hat{\Vec{z}}$ has a lower value of $f_{022} = 0$ ($f_{024} = -84 \rho^2 B^2$) than other nematic states.
Actually, UN, $D_4$BN and $D_2$BN states are predicted to be realized within the GL expansion in Eq. \eqref{eq:neutron_Free_energy} \cite{Yasui:2019unp}.

\section{Summary and Discussion}\label{sec:summary} 

In this paper, we have discussed a relationship between the neutron $^3P_2$ superfluids and spin-2 spinor BECs, the formalism of which is usually written with the $3\times 3$ traceless symmetric tensor $A$ and the five-component 
condensate wave functions $\psi$, respectively. 
Because $A$ and $\psi$ have the same internal degrees of freedom, they are transformed to each other, and we have shown the correspondence between the free-energy density $f$ for the neutron superfluids and the low-energy density $h$ for spin-2 spinor BECs in the both languages of $A$ and $\psi$.
We also have listed 15 uniform states having the internal symmetry as the candidate of the possible ground states for neutron $^3P_2$ superfluids.

In this paper, we have started from the low-energy functional $h$ in Eq.~\eqref{eq:spinor-Hamiltonian-original} and the GL free energy density $f$ in Eq.~\eqref{eq:neutron_Free_energy} which are effective theory restricted at $T = 0$ and $T \approx T_{\rm c}$, respectively. 
We need to use a different framework beyond these temperature regions.
For example, we have to use the BdG theory \cite{Mizushima:2016fbn} for neutron $^3P_2$ superfluidity near $T = 0$ 
(see Ref.~\cite{Zverev:2003ak} for $^3P_2$-$^3F_2$ pairing). 
Once we find new candidates for stable states, 
our results will immediately give information for such new states via Table \ref{table:terms}.

We hope that our present work will lead a deeper understanding of neutron $^3P_2$ superfluids, spin-2 spinor BECs, their relationship, and a systematic way to discuss possible ground states.
Although not all states appear as the ground states 
or metastable states
in either spin-2 spinor BEC or $^3P_2$ superfluids,
some may appear for instance as metastable states
at spin-2 spinor BECs with higher order interaction terms such as three- and four-body scattering,
or higher order expansions 
of the GL theory for the $^3P_2$ superfluids.
Another application of our present work is local structures in a vortex core at which the different symmetric state may appear from that surrounding the vortex in the bulk.
We can discuss which state is filled at the vortex core by calculating $\Vec{S}^2$, $|\Psi_{20}|^2$, $|\Psi_{30}|^2$, $|\Phi_{30}|^2$, and so on.
We will report on this topic elsewhere.

In Ref.~\cite{Yasui:2020xqb} a mixture of a $^3P_2$ neturon superfluid
and a $^1S_0$ neutron superfluid, which may be realized in the intermediate region of neutron star cores, was discussed. 
This situation corresponds to a mixture of a spin-2 spinor 
BEC and a scalar BEC.

\section*{Acknowledgment}

We would like to thank Shigehiro Yasui for helpful discussions and comments.
The work of M.K. is partly supported by JSPS KAKENHI (Grants No. 20K03765 and 19KK0066), and by the Osaka City University Advanced Mathematical Institute (MEXT Joint Usage/Research Center on
Mathematics and Theoretical Physics, Grant No. JPMXP0619217849).
The work of M.N. is supported in part by JSPS KAKENHI (Grant No.~JP18H01217).

\newpage

\bibliography{relation_spinorBEC}

\begin{thebibliography}{87}%
\makeatletter
\providecommand \@ifxundefined [1]{%
 \@ifx{#1\undefined}
}%
\providecommand \@ifnum [1]{%
 \ifnum #1\expandafter \@firstoftwo
 \else \expandafter \@secondoftwo
 \fi
}%
\providecommand \@ifx [1]{%
 \ifx #1\expandafter \@firstoftwo
 \else \expandafter \@secondoftwo
 \fi
}%
\providecommand \natexlab [1]{#1}%
\providecommand \enquote  [1]{``#1''}%
\providecommand \bibnamefont  [1]{#1}%
\providecommand \bibfnamefont [1]{#1}%
\providecommand \citenamefont [1]{#1}%
\providecommand \href@noop [0]{\@secondoftwo}%
\providecommand \href [0]{\begingroup \@sanitize@url \@href}%
\providecommand \@href[1]{\@@startlink{#1}\@@href}%
\providecommand \@@href[1]{\endgroup#1\@@endlink}%
\providecommand \@sanitize@url [0]{\catcode `\\12\catcode `\$12\catcode
  `\&12\catcode `\#12\catcode `\^12\catcode `\_12\catcode `\%12\relax}%
\providecommand \@@startlink[1]{}%
\providecommand \@@endlink[0]{}%
\providecommand \url  [0]{\begingroup\@sanitize@url \@url }%
\providecommand \@url [1]{\endgroup\@href {#1}{\urlprefix }}%
\providecommand \urlprefix  [0]{URL }%
\providecommand \Eprint [0]{\href }%
\providecommand \doibase [0]{https://doi.org/}%
\providecommand \selectlanguage [0]{\@gobble}%
\providecommand \bibinfo  [0]{\@secondoftwo}%
\providecommand \bibfield  [0]{\@secondoftwo}%
\providecommand \translation [1]{[#1]}%
\providecommand \BibitemOpen [0]{}%
\providecommand \bibitemStop [0]{}%
\providecommand \bibitemNoStop [0]{.\EOS\space}%
\providecommand \EOS [0]{\spacefactor3000\relax}%
\providecommand \BibitemShut  [1]{\csname bibitem#1\endcsname}%
\let\auto@bib@innerbib\@empty
%</preamble>
\bibitem [{\citenamefont {Vollhardt}\ and\ \citenamefont
  {W{\"o}lfle}(2013)}]{vollhardt2013superfluid}%
  \BibitemOpen
  \bibfield  {author} {\bibinfo {author} {\bibfnamefont {D.}~\bibnamefont
  {Vollhardt}}\ and\ \bibinfo {author} {\bibfnamefont {P.}~\bibnamefont
  {W{\"o}lfle}},\ }\href {https://books.google.co.jp/books?id=jY6yAAAAQBAJ}
  {\emph {\bibinfo {title} {The Superfluid Phases of Helium 3}}},\ Dover Books
  on Physics Series\ (\bibinfo  {publisher} {Dover Publications, New York},\
  \bibinfo {year} {2013})\BibitemShut {NoStop}%
\bibitem [{vol()}]{volovik}%
  \BibitemOpen
  \href@noop {} {}\bibinfo {note} {G. E. Volovik, {\it The Universe in a Helium
  Droplet} (Clarendon, Oxford, 2003)}\BibitemShut {NoStop}%
\bibitem [{\citenamefont {Mermin}(1974)}]{Mermin:1974zz}%
  \BibitemOpen
  \bibfield  {author} {\bibinfo {author} {\bibfnamefont {N.~D.}\ \bibnamefont
  {Mermin}},\ }\href {https://doi.org/10.1103/PhysRevA.9.868} {\bibfield
  {journal} {\bibinfo  {journal} {Phys. Rev. A}\ }\textbf {\bibinfo {volume}
  {9}},\ \bibinfo {pages} {868} (\bibinfo {year} {1974})}\BibitemShut {NoStop}%
\bibitem [{\citenamefont {{Kawaguchi}}\ and\ \citenamefont
  {{Ueda}}(2012)}]{2010arXiv1001.2072K}%
  \BibitemOpen
  \bibfield  {author} {\bibinfo {author} {\bibfnamefont {Y.}~\bibnamefont
  {{Kawaguchi}}}\ and\ \bibinfo {author} {\bibfnamefont {M.}~\bibnamefont
  {{Ueda}}},\ }\href {https://doi.org/10.1016/j.physrep.2012.07.005} {\bibfield
   {journal} {\bibinfo  {journal} {Phys. Rep.}\ }\textbf {\bibinfo {volume}
  {520}},\ \bibinfo {pages} {253 } (\bibinfo {year} {2012})},\ \Eprint
  {https://arxiv.org/abs/1001.2072} {arXiv:1001.2072 [cond-mat.quant-gas]}
  \BibitemShut {NoStop}%
\bibitem [{\citenamefont {Schmaljohann}\ \emph {et~al.}(2004)\citenamefont
  {Schmaljohann}, \citenamefont {Erhard}, \citenamefont {Kronj\"ager},
  \citenamefont {Kottke}, \citenamefont {van Staa}, \citenamefont
  {Cacciapuoti}, \citenamefont {Arlt}, \citenamefont {Bongs},\ and\
  \citenamefont {Sengstock}}]{Schmaljohann:2004}%
  \BibitemOpen
  \bibfield  {author} {\bibinfo {author} {\bibfnamefont {H.}~\bibnamefont
  {Schmaljohann}}, \bibinfo {author} {\bibfnamefont {M.}~\bibnamefont
  {Erhard}}, \bibinfo {author} {\bibfnamefont {J.}~\bibnamefont {Kronj\"ager}},
  \bibinfo {author} {\bibfnamefont {M.}~\bibnamefont {Kottke}}, \bibinfo
  {author} {\bibfnamefont {S.}~\bibnamefont {van Staa}}, \bibinfo {author}
  {\bibfnamefont {L.}~\bibnamefont {Cacciapuoti}}, \bibinfo {author}
  {\bibfnamefont {J.~J.}\ \bibnamefont {Arlt}}, \bibinfo {author}
  {\bibfnamefont {K.}~\bibnamefont {Bongs}},\ and\ \bibinfo {author}
  {\bibfnamefont {K.}~\bibnamefont {Sengstock}},\ }\href
  {https://doi.org/10.1103/PhysRevLett.92.040402} {\bibfield  {journal}
  {\bibinfo  {journal} {Phys. Rev. Lett.}\ }\textbf {\bibinfo {volume} {92}},\
  \bibinfo {pages} {040402} (\bibinfo {year} {2004})}\BibitemShut {NoStop}%
\bibitem [{\citenamefont {Chang}\ \emph {et~al.}(2004)\citenamefont {Chang},
  \citenamefont {Hamley}, \citenamefont {Barrett}, \citenamefont {Sauer},
  \citenamefont {Fortier}, \citenamefont {Zhang}, \citenamefont {You},\ and\
  \citenamefont {Chapman}}]{Chang:2004}%
  \BibitemOpen
  \bibfield  {author} {\bibinfo {author} {\bibfnamefont {M.-S.}\ \bibnamefont
  {Chang}}, \bibinfo {author} {\bibfnamefont {C.~D.}\ \bibnamefont {Hamley}},
  \bibinfo {author} {\bibfnamefont {M.~D.}\ \bibnamefont {Barrett}}, \bibinfo
  {author} {\bibfnamefont {J.~A.}\ \bibnamefont {Sauer}}, \bibinfo {author}
  {\bibfnamefont {K.~M.}\ \bibnamefont {Fortier}}, \bibinfo {author}
  {\bibfnamefont {W.}~\bibnamefont {Zhang}}, \bibinfo {author} {\bibfnamefont
  {L.}~\bibnamefont {You}},\ and\ \bibinfo {author} {\bibfnamefont {M.~S.}\
  \bibnamefont {Chapman}},\ }\href
  {https://doi.org/10.1103/PhysRevLett.92.140403} {\bibfield  {journal}
  {\bibinfo  {journal} {Phys. Rev. Lett.}\ }\textbf {\bibinfo {volume} {92}},\
  \bibinfo {pages} {140403} (\bibinfo {year} {2004})}\BibitemShut {NoStop}%
\bibitem [{\citenamefont {Kuwamoto}\ \emph {et~al.}(2004)\citenamefont
  {Kuwamoto}, \citenamefont {Araki}, \citenamefont {Eno},\ and\ \citenamefont
  {Hirano}}]{Kuwamoto:2004}%
  \BibitemOpen
  \bibfield  {author} {\bibinfo {author} {\bibfnamefont {T.}~\bibnamefont
  {Kuwamoto}}, \bibinfo {author} {\bibfnamefont {K.}~\bibnamefont {Araki}},
  \bibinfo {author} {\bibfnamefont {T.}~\bibnamefont {Eno}},\ and\ \bibinfo
  {author} {\bibfnamefont {T.}~\bibnamefont {Hirano}},\ }\href
  {https://doi.org/10.1103/PhysRevA.69.063604} {\bibfield  {journal} {\bibinfo
  {journal} {Phys. Rev. A}\ }\textbf {\bibinfo {volume} {69}},\ \bibinfo
  {pages} {063604} (\bibinfo {year} {2004})}\BibitemShut {NoStop}%
\bibitem [{\citenamefont {Widera}\ \emph {et~al.}(2006)\citenamefont {Widera},
  \citenamefont {Gerbier}, \citenamefont {F{\"o}lling}, \citenamefont
  {Gericke}, \citenamefont {Mandel},\ and\ \citenamefont
  {Bloch}}]{Widera:2006}%
  \BibitemOpen
  \bibfield  {author} {\bibinfo {author} {\bibfnamefont {A.}~\bibnamefont
  {Widera}}, \bibinfo {author} {\bibfnamefont {F.}~\bibnamefont {Gerbier}},
  \bibinfo {author} {\bibfnamefont {S.}~\bibnamefont {F{\"o}lling}}, \bibinfo
  {author} {\bibfnamefont {T.}~\bibnamefont {Gericke}}, \bibinfo {author}
  {\bibfnamefont {O.}~\bibnamefont {Mandel}},\ and\ \bibinfo {author}
  {\bibfnamefont {I.}~\bibnamefont {Bloch}},\ }\href
  {https://doi.org/10.1088/1367-2630/8/8/152} {\bibfield  {journal} {\bibinfo
  {journal} {New Journal of Physics}\ }\textbf {\bibinfo {volume} {8}},\
  \bibinfo {pages} {152} (\bibinfo {year} {2006})}\BibitemShut {NoStop}%
\bibitem [{\citenamefont {Tojo}\ \emph {et~al.}(2008)\citenamefont {Tojo},
  \citenamefont {Tomiyama}, \citenamefont {Iwata}, \citenamefont {Kuwamoto},\
  and\ \citenamefont {Hirano}}]{Tojo:2008}%
  \BibitemOpen
  \bibfield  {author} {\bibinfo {author} {\bibfnamefont {S.}~\bibnamefont
  {Tojo}}, \bibinfo {author} {\bibfnamefont {A.}~\bibnamefont {Tomiyama}},
  \bibinfo {author} {\bibfnamefont {M.}~\bibnamefont {Iwata}}, \bibinfo
  {author} {\bibfnamefont {T.}~\bibnamefont {Kuwamoto}},\ and\ \bibinfo
  {author} {\bibfnamefont {T.}~\bibnamefont {Hirano}},\ }\href
  {https://doi.org/10.1007/s00340-008-3224-y} {\bibfield  {journal} {\bibinfo
  {journal} {Applied Physics B}\ }\textbf {\bibinfo {volume} {93}},\ \bibinfo
  {pages} {403} (\bibinfo {year} {2008})}\BibitemShut {NoStop}%
\bibitem [{\citenamefont {Tojo}\ \emph {et~al.}(2009)\citenamefont {Tojo},
  \citenamefont {Hayashi}, \citenamefont {Tanabe}, \citenamefont {Hirano},
  \citenamefont {Kawaguchi}, \citenamefont {Saito},\ and\ \citenamefont
  {Ueda}}]{Tojo:2009}%
  \BibitemOpen
  \bibfield  {author} {\bibinfo {author} {\bibfnamefont {S.}~\bibnamefont
  {Tojo}}, \bibinfo {author} {\bibfnamefont {T.}~\bibnamefont {Hayashi}},
  \bibinfo {author} {\bibfnamefont {T.}~\bibnamefont {Tanabe}}, \bibinfo
  {author} {\bibfnamefont {T.}~\bibnamefont {Hirano}}, \bibinfo {author}
  {\bibfnamefont {Y.}~\bibnamefont {Kawaguchi}}, \bibinfo {author}
  {\bibfnamefont {H.}~\bibnamefont {Saito}},\ and\ \bibinfo {author}
  {\bibfnamefont {M.}~\bibnamefont {Ueda}},\ }\href
  {https://doi.org/10.1103/PhysRevA.80.042704} {\bibfield  {journal} {\bibinfo
  {journal} {Phys. Rev. A}\ }\textbf {\bibinfo {volume} {80}},\ \bibinfo
  {pages} {042704} (\bibinfo {year} {2009})}\BibitemShut {NoStop}%
\bibitem [{Note1()}]{Note1}%
  \BibitemOpen
  \bibinfo {note} {The cyclic phase is also discussed in a superconductor \cite
  {Mizushima:2017pma}.}\BibitemShut {Stop}%
\bibitem [{\citenamefont {Kobayashi}\ \emph {et~al.}(2012)\citenamefont
  {Kobayashi}, \citenamefont {Kobayashi}, \citenamefont {Kawaguchi},
  \citenamefont {Nitta},\ and\ \citenamefont {Ueda}}]{Kobayashi:2011xb}%
  \BibitemOpen
  \bibfield  {author} {\bibinfo {author} {\bibfnamefont {S.}~\bibnamefont
  {Kobayashi}}, \bibinfo {author} {\bibfnamefont {M.}~\bibnamefont
  {Kobayashi}}, \bibinfo {author} {\bibfnamefont {Y.}~\bibnamefont
  {Kawaguchi}}, \bibinfo {author} {\bibfnamefont {M.}~\bibnamefont {Nitta}},\
  and\ \bibinfo {author} {\bibfnamefont {M.}~\bibnamefont {Ueda}},\ }\href
  {https://doi.org/10.1016/j.nuclphysb.2011.11.003} {\bibfield  {journal}
  {\bibinfo  {journal} {Nucl. Phys. B}\ }\textbf {\bibinfo {volume} {856}},\
  \bibinfo {pages} {577} (\bibinfo {year} {2012})},\ \Eprint
  {https://arxiv.org/abs/1110.1478} {arXiv:1110.1478 [math-ph]} \BibitemShut
  {NoStop}%
%%CITATION = ARXIV:1110.1478;%%
\bibitem [{\citenamefont {Semenoff}\ and\ \citenamefont
  {Zhou}(2007)}]{Semenoff:2006vv}%
  \BibitemOpen
  \bibfield  {author} {\bibinfo {author} {\bibfnamefont {G.~W.}\ \bibnamefont
  {Semenoff}}\ and\ \bibinfo {author} {\bibfnamefont {F.}~\bibnamefont
  {Zhou}},\ }\href {https://doi.org/10.1103/PhysRevLett.98.100401} {\bibfield
  {journal} {\bibinfo  {journal} {Phys. Rev. Lett.}\ }\textbf {\bibinfo
  {volume} {98}},\ \bibinfo {pages} {100401} (\bibinfo {year} {2007})},\
  \Eprint {https://arxiv.org/abs/cond-mat/0610162} {arXiv:cond-mat/0610162
  [cond-mat]} \BibitemShut {NoStop}%
%%CITATION = COND-MAT/0610162;%%
\bibitem [{\citenamefont {Kobayashi}\ \emph {et~al.}(2009)\citenamefont
  {Kobayashi}, \citenamefont {Kawaguchi}, \citenamefont {Nitta},\ and\
  \citenamefont {Ueda}}]{Kobayashi:2008pk}%
  \BibitemOpen
  \bibfield  {author} {\bibinfo {author} {\bibfnamefont {M.}~\bibnamefont
  {Kobayashi}}, \bibinfo {author} {\bibfnamefont {Y.}~\bibnamefont
  {Kawaguchi}}, \bibinfo {author} {\bibfnamefont {M.}~\bibnamefont {Nitta}},\
  and\ \bibinfo {author} {\bibfnamefont {M.}~\bibnamefont {Ueda}},\ }\href
  {https://doi.org/10.1103/PhysRevLett.103.115301} {\bibfield  {journal}
  {\bibinfo  {journal} {Phys. Rev. Lett.}\ }\textbf {\bibinfo {volume} {103}},\
  \bibinfo {pages} {115301} (\bibinfo {year} {2009})},\ \Eprint
  {https://arxiv.org/abs/0810.5441} {arXiv:0810.5441 [cond-mat.other]}
  \BibitemShut {NoStop}%
%%CITATION = ARXIV:0810.5441;%%
\bibitem [{\citenamefont {Mawson}\ \emph {et~al.}(2019)\citenamefont {Mawson},
  \citenamefont {Petersen}, \citenamefont {Slingerland},\ and\ \citenamefont
  {Simula}}]{Mawson:2019}%
  \BibitemOpen
  \bibfield  {author} {\bibinfo {author} {\bibfnamefont {T.}~\bibnamefont
  {Mawson}}, \bibinfo {author} {\bibfnamefont {T.~C.}\ \bibnamefont
  {Petersen}}, \bibinfo {author} {\bibfnamefont {J.~K.}\ \bibnamefont
  {Slingerland}},\ and\ \bibinfo {author} {\bibfnamefont {T.~P.}\ \bibnamefont
  {Simula}},\ }\href {https://doi.org/10.1103/PhysRevLett.123.140404}
  {\bibfield  {journal} {\bibinfo  {journal} {Phys. Rev. Lett.}\ }\textbf
  {\bibinfo {volume} {123}},\ \bibinfo {pages} {140404} (\bibinfo {year}
  {2019})}\BibitemShut {NoStop}%
\bibitem [{\citenamefont {Tiurev}\ \emph {et~al.}(2018)\citenamefont {Tiurev},
  \citenamefont {Ollikainen}, \citenamefont {Kuopanportti}, \citenamefont
  {Nakahara}, \citenamefont {Hall},\ and\ \citenamefont
  {M\"ott\"onen}}]{Tiurev:2017xgn}%
  \BibitemOpen
  \bibfield  {author} {\bibinfo {author} {\bibfnamefont {K.}~\bibnamefont
  {Tiurev}}, \bibinfo {author} {\bibfnamefont {T.}~\bibnamefont {Ollikainen}},
  \bibinfo {author} {\bibfnamefont {P.}~\bibnamefont {Kuopanportti}}, \bibinfo
  {author} {\bibfnamefont {M.}~\bibnamefont {Nakahara}}, \bibinfo {author}
  {\bibfnamefont {D.~S.}\ \bibnamefont {Hall}},\ and\ \bibinfo {author}
  {\bibfnamefont {M.}~\bibnamefont {M\"ott\"onen}},\ }\href
  {https://doi.org/10.1088/1367-2630/aac2a8} {\bibfield  {journal} {\bibinfo
  {journal} {New J. Phys.}\ }\textbf {\bibinfo {volume} {20}},\ \bibinfo
  {pages} {055011} (\bibinfo {year} {2018})},\ \Eprint
  {https://arxiv.org/abs/1711.01155} {arXiv:1711.01155 [cond-mat.quant-gas]}
  \BibitemShut {NoStop}%
\bibitem [{\citenamefont {Uchino}\ \emph
  {et~al.}(2010{\natexlab{a}})\citenamefont {Uchino}, \citenamefont
  {Kobayashi},\ and\ \citenamefont {Ueda}}]{Uchino:2010}%
  \BibitemOpen
  \bibfield  {author} {\bibinfo {author} {\bibfnamefont {S.}~\bibnamefont
  {Uchino}}, \bibinfo {author} {\bibfnamefont {M.}~\bibnamefont {Kobayashi}},\
  and\ \bibinfo {author} {\bibfnamefont {M.}~\bibnamefont {Ueda}},\ }\href
  {https://doi.org/10.1103/PhysRevA.81.063632} {\bibfield  {journal} {\bibinfo
  {journal} {Phys. Rev. A}\ }\textbf {\bibinfo {volume} {81}},\ \bibinfo
  {pages} {063632} (\bibinfo {year} {2010}{\natexlab{a}})}\BibitemShut
  {NoStop}%
\bibitem [{\citenamefont {Uchino}\ \emph
  {et~al.}(2010{\natexlab{b}})\citenamefont {Uchino}, \citenamefont
  {Kobayashi}, \citenamefont {Nitta},\ and\ \citenamefont
  {Ueda}}]{Uchino:2010pf}%
  \BibitemOpen
  \bibfield  {author} {\bibinfo {author} {\bibfnamefont {S.}~\bibnamefont
  {Uchino}}, \bibinfo {author} {\bibfnamefont {M.}~\bibnamefont {Kobayashi}},
  \bibinfo {author} {\bibfnamefont {M.}~\bibnamefont {Nitta}},\ and\ \bibinfo
  {author} {\bibfnamefont {M.}~\bibnamefont {Ueda}},\ }\href
  {https://doi.org/10.1103/PhysRevLett.105.230406} {\bibfield  {journal}
  {\bibinfo  {journal} {Phys. Rev. Lett.}\ }\textbf {\bibinfo {volume} {105}},\
  \bibinfo {pages} {230406} (\bibinfo {year} {2010}{\natexlab{b}})},\ \Eprint
  {https://arxiv.org/abs/1010.2864} {arXiv:1010.2864 [cond-mat.quant-gas]}
  \BibitemShut {NoStop}%
%%CITATION = ARXIV:1010.2864;%%
\bibitem [{\citenamefont {Borgh}\ and\ \citenamefont
  {Ruostekoski}(2016)}]{Borgh:2016cco}%
  \BibitemOpen
  \bibfield  {author} {\bibinfo {author} {\bibfnamefont {M.~O.}\ \bibnamefont
  {Borgh}}\ and\ \bibinfo {author} {\bibfnamefont {J.}~\bibnamefont
  {Ruostekoski}},\ }\href {https://doi.org/10.1103/PhysRevLett.117.275302}
  {\bibfield  {journal} {\bibinfo  {journal} {Phys. Rev. Lett.}\ }\textbf
  {\bibinfo {volume} {117}},\ \bibinfo {pages} {275302} (\bibinfo {year}
  {2016})},\ \bibinfo {note} {[Erratum: Phys. Rev. Lett. {\bf 118}, 129901(E)
  (2017)]}\BibitemShut {NoStop}%
\bibitem [{\citenamefont {Tabakin}(1968)}]{Tabakin:1968zz}%
  \BibitemOpen
  \bibfield  {author} {\bibinfo {author} {\bibfnamefont {F.}~\bibnamefont
  {Tabakin}},\ }\href {https://doi.org/10.1103/PhysRev.174.1208} {\bibfield
  {journal} {\bibinfo  {journal} {Phys. Rev.}\ }\textbf {\bibinfo {volume}
  {174}},\ \bibinfo {pages} {1208} (\bibinfo {year} {1968})}\BibitemShut
  {NoStop}%
%%CITATION = PHRVA,174,1208;%%
\bibitem [{\citenamefont {Hoffberg}\ \emph {et~al.}(1970)\citenamefont
  {Hoffberg}, \citenamefont {Glassgold}, \citenamefont {Richardson},\ and\
  \citenamefont {Ruderman}}]{Hoffberg:1970vqj}%
  \BibitemOpen
  \bibfield  {author} {\bibinfo {author} {\bibfnamefont {M.}~\bibnamefont
  {Hoffberg}}, \bibinfo {author} {\bibfnamefont {A.~E.}\ \bibnamefont
  {Glassgold}}, \bibinfo {author} {\bibfnamefont {R.~W.}\ \bibnamefont
  {Richardson}},\ and\ \bibinfo {author} {\bibfnamefont {M.}~\bibnamefont
  {Ruderman}},\ }\href {https://doi.org/10.1103/PhysRevLett.24.775} {\bibfield
  {journal} {\bibinfo  {journal} {Phys. Rev. Lett.}\ }\textbf {\bibinfo
  {volume} {24}},\ \bibinfo {pages} {775} (\bibinfo {year} {1970})}\BibitemShut
  {NoStop}%
%%CITATION = PRLTA,24,775;%%
\bibitem [{\citenamefont {Tamagaki}(1970)}]{Tamagaki1970}%
  \BibitemOpen
  \bibfield  {author} {\bibinfo {author} {\bibfnamefont {R.}~\bibnamefont
  {Tamagaki}},\ }\href {https://doi.org/10.1143/PTP.44.905} {\bibfield
  {journal} {\bibinfo  {journal} {Progress of Theoretical Physics}\ }\textbf
  {\bibinfo {volume} {44}},\ \bibinfo {pages} {905} (\bibinfo {year}
  {1970})}\BibitemShut {NoStop}%
\bibitem [{\citenamefont {Takatsuka}\ and\ \citenamefont
  {Tamagaki}(1971)}]{Takatsuka1971}%
  \BibitemOpen
  \bibfield  {author} {\bibinfo {author} {\bibfnamefont {T.}~\bibnamefont
  {Takatsuka}}\ and\ \bibinfo {author} {\bibfnamefont {R.}~\bibnamefont
  {Tamagaki}},\ }\href {https://doi.org/10.1143/PTP.46.114} {\bibfield
  {journal} {\bibinfo  {journal} {Progress of Theoretical Physics}\ }\textbf
  {\bibinfo {volume} {46}},\ \bibinfo {pages} {114} (\bibinfo {year}
  {1971})}\BibitemShut {NoStop}%
\bibitem [{\citenamefont {Takatsuka}(1972)}]{Takatsuka1972}%
  \BibitemOpen
  \bibfield  {author} {\bibinfo {author} {\bibfnamefont {T.}~\bibnamefont
  {Takatsuka}},\ }\href {https://doi.org/10.1143/PTP.47.1062} {\bibfield
  {journal} {\bibinfo  {journal} {Progress of Theoretical Physics}\ }\textbf
  {\bibinfo {volume} {47}},\ \bibinfo {pages} {1062} (\bibinfo {year}
  {1972})}\BibitemShut {NoStop}%
\bibitem [{\citenamefont {Fujita}\ and\ \citenamefont
  {Tsuneto}(1972)}]{Fujita1972}%
  \BibitemOpen
  \bibfield  {author} {\bibinfo {author} {\bibfnamefont {T.}~\bibnamefont
  {Fujita}}\ and\ \bibinfo {author} {\bibfnamefont {T.}~\bibnamefont
  {Tsuneto}},\ }\href {https://doi.org/10.1143/PTP.48.766} {\bibfield
  {journal} {\bibinfo  {journal} {Prog. Theor. Phys.}\ }\textbf {\bibinfo
  {volume} {48}},\ \bibinfo {pages} {766} (\bibinfo {year} {1972})}\BibitemShut
  {NoStop}%
\bibitem [{\citenamefont {Richardson}(1972)}]{Richardson:1972xn}%
  \BibitemOpen
  \bibfield  {author} {\bibinfo {author} {\bibfnamefont {R.~W.}\ \bibnamefont
  {Richardson}},\ }\href {https://doi.org/10.1103/PhysRevD.5.1883} {\bibfield
  {journal} {\bibinfo  {journal} {Phys. Rev. D}\ }\textbf {\bibinfo {volume}
  {5}},\ \bibinfo {pages} {1883} (\bibinfo {year} {1972})}\BibitemShut
  {NoStop}%
%%CITATION = PHRVA,D5,1883;%%
\bibitem [{\citenamefont {Amundsen}\ and\ \citenamefont
  {Ostgaard}(1985)}]{Amundsen:1984qc}%
  \BibitemOpen
  \bibfield  {author} {\bibinfo {author} {\bibfnamefont {L.}~\bibnamefont
  {Amundsen}}\ and\ \bibinfo {author} {\bibfnamefont {E.}~\bibnamefont
  {Ostgaard}},\ }\href {https://doi.org/10.1016/0375-9474(85)90140-X,
  10.1016/S0375-9474(85)80012-9} {\bibfield  {journal} {\bibinfo  {journal}
  {Nucl. Phys. A}\ }\textbf {\bibinfo {volume} {442}},\ \bibinfo {pages} {163}
  (\bibinfo {year} {1985})}\BibitemShut {NoStop}%
%%CITATION = NUPHA,A442,163;%%
\bibitem [{\citenamefont {Takatsuka}\ and\ \citenamefont
  {Tamagaki}(1993)}]{Takatsuka:1992ga}%
  \BibitemOpen
  \bibfield  {author} {\bibinfo {author} {\bibfnamefont {T.}~\bibnamefont
  {Takatsuka}}\ and\ \bibinfo {author} {\bibfnamefont {R.}~\bibnamefont
  {Tamagaki}},\ }\href {https://doi.org/10.1143/PTPS.112.27} {\bibfield
  {journal} {\bibinfo  {journal} {Prog. Theor. Phys. Suppl.}\ }\textbf
  {\bibinfo {volume} {112}},\ \bibinfo {pages} {27} (\bibinfo {year}
  {1993})}\BibitemShut {NoStop}%
%%CITATION = PTPSA,112,27;%%
\bibitem [{\citenamefont {Baldo}\ \emph {et~al.}(1992)\citenamefont {Baldo},
  \citenamefont {Cugnon}, \citenamefont {Lejeune},\ and\ \citenamefont
  {Lombardo}}]{Baldo:1992kzz}%
  \BibitemOpen
  \bibfield  {author} {\bibinfo {author} {\bibfnamefont {M.}~\bibnamefont
  {Baldo}}, \bibinfo {author} {\bibfnamefont {J.}~\bibnamefont {Cugnon}},
  \bibinfo {author} {\bibfnamefont {A.}~\bibnamefont {Lejeune}},\ and\ \bibinfo
  {author} {\bibfnamefont {U.}~\bibnamefont {Lombardo}},\ }\href
  {https://doi.org/10.1016/0375-9474(92)90387-Y} {\bibfield  {journal}
  {\bibinfo  {journal} {Nucl. Phys. A}\ }\textbf {\bibinfo {volume} {536}},\
  \bibinfo {pages} {349} (\bibinfo {year} {1992})}\BibitemShut {NoStop}%
%%CITATION = NUPHA,A536,349;%%
\bibitem [{\citenamefont {Elgaroy}\ \emph {et~al.}(1996)\citenamefont
  {Elgaroy}, \citenamefont {Engvik}, \citenamefont {Hjorth-Jensen},\ and\
  \citenamefont {Osnes}}]{Elgaroy:1996hp}%
  \BibitemOpen
  \bibfield  {author} {\bibinfo {author} {\bibfnamefont {O.}~\bibnamefont
  {Elgaroy}}, \bibinfo {author} {\bibfnamefont {L.}~\bibnamefont {Engvik}},
  \bibinfo {author} {\bibfnamefont {M.}~\bibnamefont {Hjorth-Jensen}},\ and\
  \bibinfo {author} {\bibfnamefont {E.}~\bibnamefont {Osnes}},\ }\href
  {https://doi.org/10.1016/0375-9474(96)00217-5} {\bibfield  {journal}
  {\bibinfo  {journal} {Nucl. Phys. A}\ }\textbf {\bibinfo {volume} {607}},\
  \bibinfo {pages} {425} (\bibinfo {year} {1996})},\ \Eprint
  {https://arxiv.org/abs/nucl-th/9604032} {arXiv:nucl-th/9604032 [nucl-th]}
  \BibitemShut {NoStop}%
%%CITATION = NUCL-TH/9604032;%%
\bibitem [{\citenamefont {Khodel}\ \emph {et~al.}(1998)\citenamefont {Khodel},
  \citenamefont {Khodel},\ and\ \citenamefont {Clark}}]{Khodel:1998hn}%
  \BibitemOpen
  \bibfield  {author} {\bibinfo {author} {\bibfnamefont {V.~A.}\ \bibnamefont
  {Khodel}}, \bibinfo {author} {\bibfnamefont {V.~V.}\ \bibnamefont {Khodel}},\
  and\ \bibinfo {author} {\bibfnamefont {J.~W.}\ \bibnamefont {Clark}},\ }\href
  {https://doi.org/10.1103/PhysRevLett.81.3828} {\bibfield  {journal} {\bibinfo
   {journal} {Phys. Rev. Lett.}\ }\textbf {\bibinfo {volume} {81}},\ \bibinfo
  {pages} {3828} (\bibinfo {year} {1998})},\ \Eprint
  {https://arxiv.org/abs/nucl-th/9807034} {arXiv:nucl-th/9807034 [nucl-th]}
  \BibitemShut {NoStop}%
%%CITATION = NUCL-TH/9807034;%%
\bibitem [{\citenamefont {Baldo}\ \emph {et~al.}(1998)\citenamefont {Baldo},
  \citenamefont {Elgaroy}, \citenamefont {Engvik}, \citenamefont
  {Hjorth-Jensen},\ and\ \citenamefont {Schulze}}]{Baldo:1998ca}%
  \BibitemOpen
  \bibfield  {author} {\bibinfo {author} {\bibfnamefont {M.}~\bibnamefont
  {Baldo}}, \bibinfo {author} {\bibfnamefont {O.}~\bibnamefont {Elgaroy}},
  \bibinfo {author} {\bibfnamefont {L.}~\bibnamefont {Engvik}}, \bibinfo
  {author} {\bibfnamefont {M.}~\bibnamefont {Hjorth-Jensen}},\ and\ \bibinfo
  {author} {\bibfnamefont {H.~J.}\ \bibnamefont {Schulze}},\ }\href
  {https://doi.org/10.1103/PhysRevC.58.1921} {\bibfield  {journal} {\bibinfo
  {journal} {Phys. Rev. C}\ }\textbf {\bibinfo {volume} {58}},\ \bibinfo
  {pages} {1921} (\bibinfo {year} {1998})},\ \Eprint
  {https://arxiv.org/abs/nucl-th/9806097} {arXiv:nucl-th/9806097 [nucl-th]}
  \BibitemShut {NoStop}%
%%CITATION = NUCL-TH/9806097;%%
\bibitem [{\citenamefont {Khodel}\ \emph {et~al.}(2001)\citenamefont {Khodel},
  \citenamefont {Khodel},\ and\ \citenamefont {Clark}}]{Khodel:2000qw}%
  \BibitemOpen
  \bibfield  {author} {\bibinfo {author} {\bibfnamefont {V.~V.}\ \bibnamefont
  {Khodel}}, \bibinfo {author} {\bibfnamefont {V.~A.}\ \bibnamefont {Khodel}},\
  and\ \bibinfo {author} {\bibfnamefont {J.~W.}\ \bibnamefont {Clark}},\ }\href
  {https://doi.org/10.1016/S0375-9474(00)00351-1} {\bibfield  {journal}
  {\bibinfo  {journal} {Nucl. Phys. A}\ }\textbf {\bibinfo {volume} {679}},\
  \bibinfo {pages} {827} (\bibinfo {year} {2001})},\ \Eprint
  {https://arxiv.org/abs/nucl-th/0001006} {arXiv:nucl-th/0001006 [nucl-th]}
  \BibitemShut {NoStop}%
%%CITATION = NUCL-TH/0001006;%%
\bibitem [{\citenamefont {Zverev}\ \emph {et~al.}(2003)\citenamefont {Zverev},
  \citenamefont {Clark},\ and\ \citenamefont {Khodel}}]{Zverev:2003ak}%
  \BibitemOpen
  \bibfield  {author} {\bibinfo {author} {\bibfnamefont {M.~V.}\ \bibnamefont
  {Zverev}}, \bibinfo {author} {\bibfnamefont {J.~W.}\ \bibnamefont {Clark}},\
  and\ \bibinfo {author} {\bibfnamefont {V.~A.}\ \bibnamefont {Khodel}},\
  }\href {https://doi.org/10.1016/S0375-9474(03)00653-5} {\bibfield  {journal}
  {\bibinfo  {journal} {Nucl. Phys. A}\ }\textbf {\bibinfo {volume} {720}},\
  \bibinfo {pages} {20} (\bibinfo {year} {2003})},\ \Eprint
  {https://arxiv.org/abs/nucl-th/0301028} {arXiv:nucl-th/0301028 [nucl-th]}
  \BibitemShut {NoStop}%
%%CITATION = NUCL-TH/0301028;%%
\bibitem [{\citenamefont {Chamel}\ and\ \citenamefont
  {Haensel}(2008)}]{Chamel:2008ca}%
  \BibitemOpen
  \bibfield  {author} {\bibinfo {author} {\bibfnamefont {N.}~\bibnamefont
  {Chamel}}\ and\ \bibinfo {author} {\bibfnamefont {P.}~\bibnamefont
  {Haensel}},\ }\href {https://doi.org/10.12942/lrr-2008-10} {\bibfield
  {journal} {\bibinfo  {journal} {Living Rev. Rel.}\ }\textbf {\bibinfo
  {volume} {11}},\ \bibinfo {pages} {10} (\bibinfo {year} {2008})},\ \Eprint
  {https://arxiv.org/abs/0812.3955} {arXiv:0812.3955 [astro-ph]} \BibitemShut
  {NoStop}%
%%CITATION = ARXIV:0812.3955;%%
\bibitem [{\citenamefont {Maurizio}\ \emph {et~al.}(2014)\citenamefont
  {Maurizio}, \citenamefont {Holt},\ and\ \citenamefont
  {Finelli}}]{Maurizio:2014qsa}%
  \BibitemOpen
  \bibfield  {author} {\bibinfo {author} {\bibfnamefont {S.}~\bibnamefont
  {Maurizio}}, \bibinfo {author} {\bibfnamefont {J.~W.}\ \bibnamefont {Holt}},\
  and\ \bibinfo {author} {\bibfnamefont {P.}~\bibnamefont {Finelli}},\ }\href
  {https://doi.org/10.1103/PhysRevC.90.044003} {\bibfield  {journal} {\bibinfo
  {journal} {Phys. Rev.}\ }\textbf {\bibinfo {volume} {C90}},\ \bibinfo {pages}
  {044003} (\bibinfo {year} {2014})},\ \Eprint
  {https://arxiv.org/abs/1408.6281} {arXiv:1408.6281 [nucl-th]} \BibitemShut
  {NoStop}%
%%CITATION = ARXIV:1408.6281;%%
\bibitem [{\citenamefont {Bogner}\ \emph {et~al.}(2010)\citenamefont {Bogner},
  \citenamefont {Furnstahl},\ and\ \citenamefont {Schwenk}}]{Bogner:2009bt}%
  \BibitemOpen
  \bibfield  {author} {\bibinfo {author} {\bibfnamefont {S.~K.}\ \bibnamefont
  {Bogner}}, \bibinfo {author} {\bibfnamefont {R.~J.}\ \bibnamefont
  {Furnstahl}},\ and\ \bibinfo {author} {\bibfnamefont {A.}~\bibnamefont
  {Schwenk}},\ }\href {https://doi.org/10.1016/j.ppnp.2010.03.001} {\bibfield
  {journal} {\bibinfo  {journal} {Prog. Part. Nucl. Phys.}\ }\textbf {\bibinfo
  {volume} {65}},\ \bibinfo {pages} {94} (\bibinfo {year} {2010})},\ \Eprint
  {https://arxiv.org/abs/0912.3688} {arXiv:0912.3688 [nucl-th]} \BibitemShut
  {NoStop}%
%%CITATION = ARXIV:0912.3688;%%
\bibitem [{\citenamefont {Srinivas}\ and\ \citenamefont
  {Ramanan}(2016)}]{Srinivas:2016kir}%
  \BibitemOpen
  \bibfield  {author} {\bibinfo {author} {\bibfnamefont {S.}~\bibnamefont
  {Srinivas}}\ and\ \bibinfo {author} {\bibfnamefont {S.}~\bibnamefont
  {Ramanan}},\ }\href {https://doi.org/10.1103/PhysRevC.94.064303} {\bibfield
  {journal} {\bibinfo  {journal} {Phys. Rev. C}\ }\textbf {\bibinfo {volume}
  {94}},\ \bibinfo {pages} {064303} (\bibinfo {year} {2016})},\ \Eprint
  {https://arxiv.org/abs/1606.09053} {arXiv:1606.09053 [nucl-th]} \BibitemShut
  {NoStop}%
%%CITATION = ARXIV:1606.09053;%%
\bibitem [{\citenamefont {Haskell}\ and\ \citenamefont
  {Sedrakian}(2018)}]{Haskell:2017lkl}%
  \BibitemOpen
  \bibfield  {author} {\bibinfo {author} {\bibfnamefont {B.}~\bibnamefont
  {Haskell}}\ and\ \bibinfo {author} {\bibfnamefont {A.}~\bibnamefont
  {Sedrakian}},\ }\href {https://doi.org/10.1007/978-3-319-97616-7_8}
  {\bibfield  {journal} {\bibinfo  {journal} {Astrophys. Space Sci. Libr.}\
  }\textbf {\bibinfo {volume} {457}},\ \bibinfo {pages} {401} (\bibinfo {year}
  {2018})},\ \Eprint {https://arxiv.org/abs/1709.10340} {arXiv:1709.10340
  [astro-ph.HE]} \BibitemShut {NoStop}%
%%CITATION = ARXIV:1709.10340;%%
\bibitem [{\citenamefont {Chamel}(2017)}]{Chamel2017}%
  \BibitemOpen
  \bibfield  {author} {\bibinfo {author} {\bibfnamefont {N.}~\bibnamefont
  {Chamel}},\ }\href {https://doi.org/10.1007/s12036-017-9470-9} {\bibfield
  {journal} {\bibinfo  {journal} {J. Astrophys. Astron.}\ }\textbf {\bibinfo
  {volume} {38}},\ \bibinfo {pages} {43} (\bibinfo {year} {2017})}\BibitemShut
  {NoStop}%
\bibitem [{\citenamefont {Sedrakian}\ and\ \citenamefont
  {Clark}(2019)}]{Sedrakian:2018ydt}%
  \BibitemOpen
  \bibfield  {author} {\bibinfo {author} {\bibfnamefont {A.}~\bibnamefont
  {Sedrakian}}\ and\ \bibinfo {author} {\bibfnamefont {J.~W.}\ \bibnamefont
  {Clark}},\ }\href {https://doi.org/10.1140/epja/i2019-12863-6} {\bibfield
  {journal} {\bibinfo  {journal} {Eur. Phys. J. A}\ }\textbf {\bibinfo {volume}
  {55}},\ \bibinfo {pages} {167} (\bibinfo {year} {2019})},\ \Eprint
  {https://arxiv.org/abs/1802.00017} {arXiv:1802.00017 [nucl-th]} \BibitemShut
  {NoStop}%
%%CITATION = ARXIV:1802.00017;%%
\bibitem [{\citenamefont {Graber}\ \emph {et~al.}(2017)\citenamefont {Graber},
  \citenamefont {Andersson},\ and\ \citenamefont {Hogg}}]{Graber:2016imq}%
  \BibitemOpen
  \bibfield  {author} {\bibinfo {author} {\bibfnamefont {V.}~\bibnamefont
  {Graber}}, \bibinfo {author} {\bibfnamefont {N.}~\bibnamefont {Andersson}},\
  and\ \bibinfo {author} {\bibfnamefont {M.}~\bibnamefont {Hogg}},\ }\href
  {https://doi.org/10.1142/S0218271817300154} {\bibfield  {journal} {\bibinfo
  {journal} {Int. J. Mod. Phys. D}\ }\textbf {\bibinfo {volume} {26}},\
  \bibinfo {pages} {1730015} (\bibinfo {year} {2017})},\ \Eprint
  {https://arxiv.org/abs/1610.06882} {arXiv:1610.06882 [astro-ph.HE]}
  \BibitemShut {NoStop}%
%%CITATION = ARXIV:1610.06882;%%
\bibitem [{\citenamefont {Andersson}(2021)}]{Andersson2021}%
  \BibitemOpen
  \bibfield  {author} {\bibinfo {author} {\bibfnamefont {N.}~\bibnamefont
  {Andersson}},\ }\bibfield  {journal} {\bibinfo  {journal} {Universe}\
  }\textbf {\bibinfo {volume} {7}},\ \href
  {https://doi.org/10.3390/universe7010017} {10.3390/universe7010017} (\bibinfo
  {year} {2021})\BibitemShut {NoStop}%
\bibitem [{\citenamefont {Heinke}\ and\ \citenamefont
  {Ho}(2010)}]{Heinke:2010cr}%
  \BibitemOpen
  \bibfield  {author} {\bibinfo {author} {\bibfnamefont {C.~O.}\ \bibnamefont
  {Heinke}}\ and\ \bibinfo {author} {\bibfnamefont {W.~C.~G.}\ \bibnamefont
  {Ho}},\ }\href {https://doi.org/10.1088/2041-8205/719/2/L167} {\bibfield
  {journal} {\bibinfo  {journal} {Astrophys. J.}\ }\textbf {\bibinfo {volume}
  {719}},\ \bibinfo {pages} {L167} (\bibinfo {year} {2010})},\ \Eprint
  {https://arxiv.org/abs/1007.4719} {arXiv:1007.4719 [astro-ph.HE]}
  \BibitemShut {NoStop}%
%%CITATION = ARXIV:1007.4719;%%
\bibitem [{\citenamefont {Shternin}\ \emph {et~al.}(2011)\citenamefont
  {Shternin}, \citenamefont {Yakovlev}, \citenamefont {Heinke}, \citenamefont
  {Ho},\ and\ \citenamefont {Patnaude}}]{Shternin2011}%
  \BibitemOpen
  \bibfield  {author} {\bibinfo {author} {\bibfnamefont {P.~S.}\ \bibnamefont
  {Shternin}}, \bibinfo {author} {\bibfnamefont {D.~G.}\ \bibnamefont
  {Yakovlev}}, \bibinfo {author} {\bibfnamefont {C.~O.}\ \bibnamefont
  {Heinke}}, \bibinfo {author} {\bibfnamefont {W.~C.~G.}\ \bibnamefont {Ho}},\
  and\ \bibinfo {author} {\bibfnamefont {D.~J.}\ \bibnamefont {Patnaude}},\
  }\href {https://doi.org/10.1111/j.1745-3933.2011.01015.x} {\bibfield
  {journal} {\bibinfo  {journal} {Mon. Not. Roy. Astron. Soc. Lett.}\ }\textbf
  {\bibinfo {volume} {412}},\ \bibinfo {pages} {L108} (\bibinfo {year}
  {2011})}\BibitemShut {NoStop}%
\bibitem [{\citenamefont {Page}\ \emph {et~al.}(2011)\citenamefont {Page},
  \citenamefont {Prakash}, \citenamefont {Lattimer},\ and\ \citenamefont
  {Steiner}}]{Page:2010aw}%
  \BibitemOpen
  \bibfield  {author} {\bibinfo {author} {\bibfnamefont {D.}~\bibnamefont
  {Page}}, \bibinfo {author} {\bibfnamefont {M.}~\bibnamefont {Prakash}},
  \bibinfo {author} {\bibfnamefont {J.~M.}\ \bibnamefont {Lattimer}},\ and\
  \bibinfo {author} {\bibfnamefont {A.~W.}\ \bibnamefont {Steiner}},\ }\href
  {https://doi.org/10.1103/PhysRevLett.106.081101} {\bibfield  {journal}
  {\bibinfo  {journal} {Phys. Rev. Lett.}\ }\textbf {\bibinfo {volume} {106}},\
  \bibinfo {pages} {081101} (\bibinfo {year} {2011})},\ \Eprint
  {https://arxiv.org/abs/1011.6142} {arXiv:1011.6142 [astro-ph.HE]}
  \BibitemShut {NoStop}%
%%CITATION = ARXIV:1011.6142;%%
\bibitem [{\citenamefont {Sauls}\ and\ \citenamefont
  {Serene}(1978)}]{Sauls:1978lna}%
  \BibitemOpen
  \bibfield  {author} {\bibinfo {author} {\bibfnamefont {J.~A.}\ \bibnamefont
  {Sauls}}\ and\ \bibinfo {author} {\bibfnamefont {J.~W.}\ \bibnamefont
  {Serene}},\ }\href {https://doi.org/10.1103/PhysRevD.17.1524} {\bibfield
  {journal} {\bibinfo  {journal} {Phys. Rev. D}\ }\textbf {\bibinfo {volume}
  {17}},\ \bibinfo {pages} {1524} (\bibinfo {year} {1978})}\BibitemShut
  {NoStop}%
%%CITATION = PHRVA,D17,1524;%%
\bibitem [{\citenamefont {Muzikar}\ \emph {et~al.}(1980)\citenamefont
  {Muzikar}, \citenamefont {Sauls},\ and\ \citenamefont
  {Serene}}]{Muzikar:1980as}%
  \BibitemOpen
  \bibfield  {author} {\bibinfo {author} {\bibfnamefont {P.}~\bibnamefont
  {Muzikar}}, \bibinfo {author} {\bibfnamefont {J.~A.}\ \bibnamefont {Sauls}},\
  and\ \bibinfo {author} {\bibfnamefont {J.~W.}\ \bibnamefont {Serene}},\
  }\href {https://doi.org/10.1103/PhysRevD.21.1494} {\bibfield  {journal}
  {\bibinfo  {journal} {Phys. Rev. D}\ }\textbf {\bibinfo {volume} {21}},\
  \bibinfo {pages} {1494} (\bibinfo {year} {1980})}\BibitemShut {NoStop}%
%%CITATION = PHRVA,D21,1494;%%
\bibitem [{\citenamefont {Sauls}\ \emph {et~al.}(1982)\citenamefont {Sauls},
  \citenamefont {Stein},\ and\ \citenamefont {Serene}}]{Sauls:1982ie}%
  \BibitemOpen
  \bibfield  {author} {\bibinfo {author} {\bibfnamefont {J.~A.}\ \bibnamefont
  {Sauls}}, \bibinfo {author} {\bibfnamefont {D.~L.}\ \bibnamefont {Stein}},\
  and\ \bibinfo {author} {\bibfnamefont {J.~W.}\ \bibnamefont {Serene}},\
  }\href {https://doi.org/10.1103/PhysRevD.25.967} {\bibfield  {journal}
  {\bibinfo  {journal} {Phys. Rev. D}\ }\textbf {\bibinfo {volume} {25}},\
  \bibinfo {pages} {967} (\bibinfo {year} {1982})}\BibitemShut {NoStop}%
%%CITATION = PHRVA,D25,967;%%
\bibitem [{\citenamefont {Vulovic}\ and\ \citenamefont
  {Sauls}(1984)}]{Vulovic:1984kc}%
  \BibitemOpen
  \bibfield  {author} {\bibinfo {author} {\bibfnamefont {V.~Z.}\ \bibnamefont
  {Vulovic}}\ and\ \bibinfo {author} {\bibfnamefont {J.~A.}\ \bibnamefont
  {Sauls}},\ }\href {https://doi.org/10.1103/PhysRevD.29.2705} {\bibfield
  {journal} {\bibinfo  {journal} {Phys. Rev. D}\ }\textbf {\bibinfo {volume}
  {29}},\ \bibinfo {pages} {2705} (\bibinfo {year} {1984})}\BibitemShut
  {NoStop}%
%%CITATION = PHRVA,D29,2705;%%
\bibitem [{\citenamefont {Masuda}\ and\ \citenamefont
  {Nitta}(2016)}]{Masuda:2015jka}%
  \BibitemOpen
  \bibfield  {author} {\bibinfo {author} {\bibfnamefont {K.}~\bibnamefont
  {Masuda}}\ and\ \bibinfo {author} {\bibfnamefont {M.}~\bibnamefont {Nitta}},\
  }\href {https://doi.org/10.1103/PhysRevC.93.035804} {\bibfield  {journal}
  {\bibinfo  {journal} {Phys. Rev. C}\ }\textbf {\bibinfo {volume} {93}},\
  \bibinfo {pages} {035804} (\bibinfo {year} {2016})},\ \Eprint
  {https://arxiv.org/abs/1512.01946} {arXiv:1512.01946 [nucl-th]} \BibitemShut
  {NoStop}%
%%CITATION = ARXIV:1512.01946;%%
\bibitem [{\citenamefont {Yasui}\ \emph
  {et~al.}(2019{\natexlab{a}})\citenamefont {Yasui}, \citenamefont
  {Chatterjee},\ and\ \citenamefont {Nitta}}]{Yasui:2018tcr}%
  \BibitemOpen
  \bibfield  {author} {\bibinfo {author} {\bibfnamefont {S.}~\bibnamefont
  {Yasui}}, \bibinfo {author} {\bibfnamefont {C.}~\bibnamefont {Chatterjee}},\
  and\ \bibinfo {author} {\bibfnamefont {M.}~\bibnamefont {Nitta}},\ }\href
  {https://doi.org/10.1103/PhysRevC.99.035213} {\bibfield  {journal} {\bibinfo
  {journal} {Phys. Rev. C}\ }\textbf {\bibinfo {volume} {99}},\ \bibinfo
  {pages} {035213} (\bibinfo {year} {2019}{\natexlab{a}})},\ \Eprint
  {https://arxiv.org/abs/1810.04901} {arXiv:1810.04901 [nucl-th]} \BibitemShut
  {NoStop}%
%%CITATION = ARXIV:1810.04901;%%
\bibitem [{\citenamefont {Yasui}\ \emph
  {et~al.}(2019{\natexlab{b}})\citenamefont {Yasui}, \citenamefont
  {Chatterjee},\ and\ \citenamefont {Nitta}}]{Yasui:2019tgc}%
  \BibitemOpen
  \bibfield  {author} {\bibinfo {author} {\bibfnamefont {S.}~\bibnamefont
  {Yasui}}, \bibinfo {author} {\bibfnamefont {C.}~\bibnamefont {Chatterjee}},\
  and\ \bibinfo {author} {\bibfnamefont {M.}~\bibnamefont {Nitta}},\ }\bibfield
   {booktitle} {\emph {\bibinfo {booktitle} {{Proceedings, 8th International
  Conference on Quarks and Nuclear Physics (QNP2018): Tsukuba, Japan, November
  13-17, 2018}}},\ }\href {https://doi.org/10.7566/JPSCP.26.024022} {\bibfield
  {journal} {\bibinfo  {journal} {JPS Conf. Proc.}\ }\textbf {\bibinfo {volume}
  {26}},\ \bibinfo {pages} {024022} (\bibinfo {year} {2019}{\natexlab{b}})},\
  \Eprint {https://arxiv.org/abs/1902.00674} {arXiv:1902.00674 [nucl-th]}
  \BibitemShut {NoStop}%
%%CITATION = ARXIV:1902.00674;%%
\bibitem [{\citenamefont {Yasui}\ \emph
  {et~al.}(2019{\natexlab{c}})\citenamefont {Yasui}, \citenamefont
  {Chatterjee}, \citenamefont {Kobayashi},\ and\ \citenamefont
  {Nitta}}]{Yasui:2019unp}%
  \BibitemOpen
  \bibfield  {author} {\bibinfo {author} {\bibfnamefont {S.}~\bibnamefont
  {Yasui}}, \bibinfo {author} {\bibfnamefont {C.}~\bibnamefont {Chatterjee}},
  \bibinfo {author} {\bibfnamefont {M.}~\bibnamefont {Kobayashi}},\ and\
  \bibinfo {author} {\bibfnamefont {M.}~\bibnamefont {Nitta}},\ }\href
  {https://doi.org/10.1103/PhysRevC.100.025204} {\bibfield  {journal} {\bibinfo
   {journal} {Phys. Rev. C}\ }\textbf {\bibinfo {volume} {100}},\ \bibinfo
  {pages} {025204} (\bibinfo {year} {2019}{\natexlab{c}})},\ \Eprint
  {https://arxiv.org/abs/1904.11399} {arXiv:1904.11399 [nucl-th]} \BibitemShut
  {NoStop}%
%%CITATION = ARXIV:1904.11399;%%
\bibitem [{\citenamefont {Yasui}\ \emph
  {et~al.}(2020{\natexlab{a}})\citenamefont {Yasui}, \citenamefont {Inotani},\
  and\ \citenamefont {Nitta}}]{Yasui:2020xqb}%
  \BibitemOpen
  \bibfield  {author} {\bibinfo {author} {\bibfnamefont {S.}~\bibnamefont
  {Yasui}}, \bibinfo {author} {\bibfnamefont {D.}~\bibnamefont {Inotani}},\
  and\ \bibinfo {author} {\bibfnamefont {M.}~\bibnamefont {Nitta}},\ }\href
  {https://doi.org/10.1103/PhysRevC.101.055806} {\bibfield  {journal} {\bibinfo
   {journal} {Phys. Rev. C}\ }\textbf {\bibinfo {volume} {101}},\ \bibinfo
  {pages} {055806} (\bibinfo {year} {2020}{\natexlab{a}})},\ \Eprint
  {https://arxiv.org/abs/2002.05429} {arXiv:2002.05429 [nucl-th]} \BibitemShut
  {NoStop}%
\bibitem [{\citenamefont {Bedaque}\ \emph {et~al.}(2003)\citenamefont
  {Bedaque}, \citenamefont {Rupak},\ and\ \citenamefont
  {Savage}}]{Bedaque:2003wj}%
  \BibitemOpen
  \bibfield  {author} {\bibinfo {author} {\bibfnamefont {P.~F.}\ \bibnamefont
  {Bedaque}}, \bibinfo {author} {\bibfnamefont {G.}~\bibnamefont {Rupak}},\
  and\ \bibinfo {author} {\bibfnamefont {M.~J.}\ \bibnamefont {Savage}},\
  }\href {https://doi.org/10.1103/PhysRevC.68.065802} {\bibfield  {journal}
  {\bibinfo  {journal} {Phys. Rev. C}\ }\textbf {\bibinfo {volume} {68}},\
  \bibinfo {pages} {065802} (\bibinfo {year} {2003})},\ \Eprint
  {https://arxiv.org/abs/nucl-th/0305032} {arXiv:nucl-th/0305032 [nucl-th]}
  \BibitemShut {NoStop}%
%%CITATION = NUCL-TH/0305032;%%
\bibitem [{\citenamefont {Leinson}(2011{\natexlab{a}})}]{Leinson:2011wf}%
  \BibitemOpen
  \bibfield  {author} {\bibinfo {author} {\bibfnamefont {L.~B.}\ \bibnamefont
  {Leinson}},\ }\href {https://doi.org/10.1016/j.physletb.2011.07.025}
  {\bibfield  {journal} {\bibinfo  {journal} {Phys. Lett. B}\ }\textbf
  {\bibinfo {volume} {702}},\ \bibinfo {pages} {422} (\bibinfo {year}
  {2011}{\natexlab{a}})},\ \Eprint {https://arxiv.org/abs/1107.4025}
  {arXiv:1107.4025 [nucl-th]} \BibitemShut {NoStop}%
%%CITATION = ARXIV:1107.4025;%%
\bibitem [{\citenamefont {Leinson}(2012)}]{Leinson:2012pn}%
  \BibitemOpen
  \bibfield  {author} {\bibinfo {author} {\bibfnamefont {L.~B.}\ \bibnamefont
  {Leinson}},\ }\href {https://doi.org/10.1103/PhysRevC.85.065502} {\bibfield
  {journal} {\bibinfo  {journal} {Phys. Rev. C}\ }\textbf {\bibinfo {volume}
  {85}},\ \bibinfo {pages} {065502} (\bibinfo {year} {2012})},\ \Eprint
  {https://arxiv.org/abs/1206.3648} {arXiv:1206.3648 [nucl-th]} \BibitemShut
  {NoStop}%
%%CITATION = ARXIV:1206.3648;%%
\bibitem [{\citenamefont {Leinson}(2013)}]{Leinson:2013si}%
  \BibitemOpen
  \bibfield  {author} {\bibinfo {author} {\bibfnamefont {L.~B.}\ \bibnamefont
  {Leinson}},\ }\href {https://doi.org/10.1103/PhysRevC.87.025501} {\bibfield
  {journal} {\bibinfo  {journal} {Phys. Rev. C}\ }\textbf {\bibinfo {volume}
  {87}},\ \bibinfo {pages} {025501} (\bibinfo {year} {2013})},\ \Eprint
  {https://arxiv.org/abs/1301.5439} {arXiv:1301.5439 [nucl-th]} \BibitemShut
  {NoStop}%
%%CITATION = ARXIV:1301.5439;%%
\bibitem [{\citenamefont {Bedaque}\ and\ \citenamefont
  {Nicholson}(2013)}]{Bedaque:2012bs}%
  \BibitemOpen
  \bibfield  {author} {\bibinfo {author} {\bibfnamefont {P.~F.}\ \bibnamefont
  {Bedaque}}\ and\ \bibinfo {author} {\bibfnamefont {A.~N.}\ \bibnamefont
  {Nicholson}},\ }\href {https://doi.org/10.1103/PhysRevC.89.029902,
  10.1103/PhysRevC.87.055807} {\bibfield  {journal} {\bibinfo  {journal} {Phys.
  Rev. C}\ }\textbf {\bibinfo {volume} {87}},\ \bibinfo {pages} {055807}
  (\bibinfo {year} {2013})},\ \bibinfo {note} {[Erratum: Phys. Rev. C
  \textbf{89}, 029902(E) (2014)]},\ \Eprint {https://arxiv.org/abs/1212.1122}
  {arXiv:1212.1122 [nucl-th]} \BibitemShut {NoStop}%
%%CITATION = ARXIV:1212.1122;%%
\bibitem [{\citenamefont {Bedaque}\ and\ \citenamefont
  {Sen}(2014)}]{Bedaque:2013rya}%
  \BibitemOpen
  \bibfield  {author} {\bibinfo {author} {\bibfnamefont {P.}~\bibnamefont
  {Bedaque}}\ and\ \bibinfo {author} {\bibfnamefont {S.}~\bibnamefont {Sen}},\
  }\href {https://doi.org/10.1103/PhysRevC.89.035808} {\bibfield  {journal}
  {\bibinfo  {journal} {Phys. Rev.}\ }\textbf {\bibinfo {volume} {C89}},\
  \bibinfo {pages} {035808} (\bibinfo {year} {2014})},\ \Eprint
  {https://arxiv.org/abs/1312.6632} {arXiv:1312.6632 [nucl-th]} \BibitemShut
  {NoStop}%
%%CITATION = ARXIV:1312.6632;%%
\bibitem [{\citenamefont {Bedaque}\ and\ \citenamefont
  {Reddy}(2014)}]{Bedaque:2013fja}%
  \BibitemOpen
  \bibfield  {author} {\bibinfo {author} {\bibfnamefont {P.~F.}\ \bibnamefont
  {Bedaque}}\ and\ \bibinfo {author} {\bibfnamefont {S.}~\bibnamefont
  {Reddy}},\ }\href {https://doi.org/10.1016/j.physletb.2014.06.033} {\bibfield
   {journal} {\bibinfo  {journal} {Phys. Lett. B}\ }\textbf {\bibinfo {volume}
  {735}},\ \bibinfo {pages} {340} (\bibinfo {year} {2014})},\ \Eprint
  {https://arxiv.org/abs/1307.8183} {arXiv:1307.8183 [nucl-th]} \BibitemShut
  {NoStop}%
%%CITATION = ARXIV:1307.8183;%%
\bibitem [{\citenamefont {Bedaque}\ \emph {et~al.}(2015)\citenamefont
  {Bedaque}, \citenamefont {Nicholson},\ and\ \citenamefont
  {Sen}}]{Bedaque:2014zta}%
  \BibitemOpen
  \bibfield  {author} {\bibinfo {author} {\bibfnamefont {P.~F.}\ \bibnamefont
  {Bedaque}}, \bibinfo {author} {\bibfnamefont {A.~N.}\ \bibnamefont
  {Nicholson}},\ and\ \bibinfo {author} {\bibfnamefont {S.}~\bibnamefont
  {Sen}},\ }\href {https://doi.org/10.1103/PhysRevC.92.035809} {\bibfield
  {journal} {\bibinfo  {journal} {Phys. Rev. C}\ }\textbf {\bibinfo {volume}
  {92}},\ \bibinfo {pages} {035809} (\bibinfo {year} {2015})},\ \Eprint
  {https://arxiv.org/abs/1408.5145} {arXiv:1408.5145 [nucl-th]} \BibitemShut
  {NoStop}%
%%CITATION = ARXIV:1408.5145;%%
\bibitem [{\citenamefont {Leinson}(2010{\natexlab{a}})}]{Leinson:2009nu}%
  \BibitemOpen
  \bibfield  {author} {\bibinfo {author} {\bibfnamefont {L.~B.}\ \bibnamefont
  {Leinson}},\ }\href {https://doi.org/10.1103/PhysRevC.81.025501} {\bibfield
  {journal} {\bibinfo  {journal} {Phys. Rev. C}\ }\textbf {\bibinfo {volume}
  {81}},\ \bibinfo {pages} {025501} (\bibinfo {year} {2010}{\natexlab{a}})},\
  \Eprint {https://arxiv.org/abs/0912.2164} {arXiv:0912.2164 [astro-ph.SR]}
  \BibitemShut {NoStop}%
%%CITATION = ARXIV:0912.2164;%%
\bibitem [{\citenamefont {Leinson}(2010{\natexlab{b}})}]{Leinson:2010yf}%
  \BibitemOpen
  \bibfield  {author} {\bibinfo {author} {\bibfnamefont {L.~B.}\ \bibnamefont
  {Leinson}},\ }\href {https://doi.org/10.1016/j.physletb.2010.04.046}
  {\bibfield  {journal} {\bibinfo  {journal} {Phys. Lett. B}\ }\textbf
  {\bibinfo {volume} {689}},\ \bibinfo {pages} {60} (\bibinfo {year}
  {2010}{\natexlab{b}})},\ \Eprint {https://arxiv.org/abs/1001.2617}
  {arXiv:1001.2617 [astro-ph.SR]} \BibitemShut {NoStop}%
%%CITATION = ARXIV:1001.2617;%%
\bibitem [{\citenamefont {Leinson}(2010{\natexlab{c}})}]{Leinson:2010pk}%
  \BibitemOpen
  \bibfield  {author} {\bibinfo {author} {\bibfnamefont {L.~B.}\ \bibnamefont
  {Leinson}},\ }\href {https://doi.org/10.1103/PhysRevC.82.065503,
  10.1103/PhysRevC.84.049901} {\bibfield  {journal} {\bibinfo  {journal} {Phys.
  Rev. C}\ }\textbf {\bibinfo {volume} {82}},\ \bibinfo {pages} {065503}
  (\bibinfo {year} {2010}{\natexlab{c}})},\ \bibinfo {note} {[Erratum: Phys.
  Rev.C84, 049901(E) (2011)]},\ \Eprint {https://arxiv.org/abs/1012.5387}
  {arXiv:1012.5387 [hep-ph]} \BibitemShut {NoStop}%
%%CITATION = ARXIV:1012.5387;%%
\bibitem [{\citenamefont {Leinson}(2011{\natexlab{b}})}]{Leinson:2010ru}%
  \BibitemOpen
  \bibfield  {author} {\bibinfo {author} {\bibfnamefont {L.~B.}\ \bibnamefont
  {Leinson}},\ }\href {https://doi.org/10.1103/PhysRevC.83.055803} {\bibfield
  {journal} {\bibinfo  {journal} {Phys. Rev. C}\ }\textbf {\bibinfo {volume}
  {83}},\ \bibinfo {pages} {055803} (\bibinfo {year} {2011}{\natexlab{b}})},\
  \Eprint {https://arxiv.org/abs/1007.2803} {arXiv:1007.2803 [hep-ph]}
  \BibitemShut {NoStop}%
%%CITATION = ARXIV:1007.2803;%%
\bibitem [{\citenamefont {Leinson}(2011{\natexlab{c}})}]{Leinson:2011jr}%
  \BibitemOpen
  \bibfield  {author} {\bibinfo {author} {\bibfnamefont {L.~B.}\ \bibnamefont
  {Leinson}},\ }\href {https://doi.org/10.1103/PhysRevC.84.045501} {\bibfield
  {journal} {\bibinfo  {journal} {Phys. Rev. C}\ }\textbf {\bibinfo {volume}
  {84}},\ \bibinfo {pages} {045501} (\bibinfo {year}
  {2011}{\natexlab{c}})}\BibitemShut {NoStop}%
\bibitem [{\citenamefont {Chatterjee}\ \emph {et~al.}(2017)\citenamefont
  {Chatterjee}, \citenamefont {Haberichter},\ and\ \citenamefont
  {Nitta}}]{Chatterjee:2016gpm}%
  \BibitemOpen
  \bibfield  {author} {\bibinfo {author} {\bibfnamefont {C.}~\bibnamefont
  {Chatterjee}}, \bibinfo {author} {\bibfnamefont {M.}~\bibnamefont
  {Haberichter}},\ and\ \bibinfo {author} {\bibfnamefont {M.}~\bibnamefont
  {Nitta}},\ }\href {https://doi.org/10.1103/PhysRevC.96.055807} {\bibfield
  {journal} {\bibinfo  {journal} {Phys. Rev.}\ }\textbf {\bibinfo {volume}
  {C96}},\ \bibinfo {pages} {055807} (\bibinfo {year} {2017})},\ \Eprint
  {https://arxiv.org/abs/1612.05588} {arXiv:1612.05588 [nucl-th]} \BibitemShut
  {NoStop}%
%%CITATION = ARXIV:1612.05588;%%
\bibitem [{\citenamefont {Masuda}\ and\ \citenamefont
  {Nitta}(2020)}]{Masuda:2016vak}%
  \BibitemOpen
  \bibfield  {author} {\bibinfo {author} {\bibfnamefont {K.}~\bibnamefont
  {Masuda}}\ and\ \bibinfo {author} {\bibfnamefont {M.}~\bibnamefont {Nitta}},\
  }\href {https://doi.org/10.1093/ptep/ptz138} {\bibfield  {journal} {\bibinfo
  {journal} {Prog. Theor. Exp. Phys.}\ }\textbf {\bibinfo {volume} {2020}},\
  \bibinfo {pages} {013} (\bibinfo {year} {2020})},\ \Eprint
  {https://arxiv.org/abs/1602.07050} {arXiv:1602.07050 [nucl-th]} \BibitemShut
  {NoStop}%
%%CITATION = ARXIV:1602.07050;%%
\bibitem [{\citenamefont {Leinson}(2020)}]{Leinson:2020xjz}%
  \BibitemOpen
  \bibfield  {author} {\bibinfo {author} {\bibfnamefont {L.~B.}\ \bibnamefont
  {Leinson}},\ }\href {https://doi.org/10.1093/mnras/staa2475} {\bibfield
  {journal} {\bibinfo  {journal} {Mon. Not. Roy. Astron. Soc.}\ }\textbf
  {\bibinfo {volume} {498}},\ \bibinfo {pages} {304} (\bibinfo {year}
  {2020})}\BibitemShut {NoStop}%
\bibitem [{\citenamefont {Yasui}\ and\ \citenamefont
  {Nitta}(2020)}]{Yasui:2019vci}%
  \BibitemOpen
  \bibfield  {author} {\bibinfo {author} {\bibfnamefont {S.}~\bibnamefont
  {Yasui}}\ and\ \bibinfo {author} {\bibfnamefont {M.}~\bibnamefont {Nitta}},\
  }\href {https://doi.org/10.1103/PhysRevC.101.015207} {\bibfield  {journal}
  {\bibinfo  {journal} {Phys. Rev. C}\ }\textbf {\bibinfo {volume} {101}},\
  \bibinfo {pages} {015207} (\bibinfo {year} {2020})}\BibitemShut {NoStop}%
\bibitem [{\citenamefont {Yasui}\ \emph
  {et~al.}(2020{\natexlab{b}})\citenamefont {Yasui}, \citenamefont
  {Chatterjee},\ and\ \citenamefont {Nitta}}]{Yasui:2019pgb}%
  \BibitemOpen
  \bibfield  {author} {\bibinfo {author} {\bibfnamefont {S.}~\bibnamefont
  {Yasui}}, \bibinfo {author} {\bibfnamefont {C.}~\bibnamefont {Chatterjee}},\
  and\ \bibinfo {author} {\bibfnamefont {M.}~\bibnamefont {Nitta}},\
  }\href@noop {} {\bibfield  {journal} {\bibinfo  {journal} {Phys. Rev.}\
  }\textbf {\bibinfo {volume} {C}} (\bibinfo {year} {2020}{\natexlab{b}})},\
  \Eprint {https://arxiv.org/abs/1905.13666} {arXiv:1905.13666 [nucl-th]}
  \BibitemShut {NoStop}%
%%CITATION = ARXIV:1905.13666;%%
\bibitem [{\citenamefont {Mizushima}\ \emph {et~al.}(2017)\citenamefont
  {Mizushima}, \citenamefont {Masuda},\ and\ \citenamefont
  {Nitta}}]{Mizushima:2016fbn}%
  \BibitemOpen
  \bibfield  {author} {\bibinfo {author} {\bibfnamefont {T.}~\bibnamefont
  {Mizushima}}, \bibinfo {author} {\bibfnamefont {K.}~\bibnamefont {Masuda}},\
  and\ \bibinfo {author} {\bibfnamefont {M.}~\bibnamefont {Nitta}},\ }\href
  {https://doi.org/10.1103/PhysRevB.95.140503} {\bibfield  {journal} {\bibinfo
  {journal} {Phys. Rev. B}\ }\textbf {\bibinfo {volume} {95}},\ \bibinfo
  {pages} {140503(R)} (\bibinfo {year} {2017})},\ \Eprint
  {https://arxiv.org/abs/1607.07266} {arXiv:1607.07266 [cond-mat.supr-con]}
  \BibitemShut {NoStop}%
%%CITATION = ARXIV:1607.07266;%%
\bibitem [{\citenamefont {Mizushima}\ \emph {et~al.}(2020)\citenamefont
  {Mizushima}, \citenamefont {Yasui},\ and\ \citenamefont
  {Nitta}}]{Mizushima:2019spl}%
  \BibitemOpen
  \bibfield  {author} {\bibinfo {author} {\bibfnamefont {T.}~\bibnamefont
  {Mizushima}}, \bibinfo {author} {\bibfnamefont {S.}~\bibnamefont {Yasui}},\
  and\ \bibinfo {author} {\bibfnamefont {M.}~\bibnamefont {Nitta}},\ }\href
  {https://doi.org/10.1103/PhysRevResearch.2.013194} {\bibfield  {journal}
  {\bibinfo  {journal} {Phys. Rev. Research}\ }\textbf {\bibinfo {volume}
  {2}},\ \bibinfo {pages} {013194} (\bibinfo {year} {2020})}\BibitemShut
  {NoStop}%
\bibitem [{\citenamefont {Masaki}\ \emph {et~al.}(2020)\citenamefont {Masaki},
  \citenamefont {Mizushima},\ and\ \citenamefont {Nitta}}]{Masaki:2019rsz}%
  \BibitemOpen
  \bibfield  {author} {\bibinfo {author} {\bibfnamefont {Y.}~\bibnamefont
  {Masaki}}, \bibinfo {author} {\bibfnamefont {T.}~\bibnamefont {Mizushima}},\
  and\ \bibinfo {author} {\bibfnamefont {M.}~\bibnamefont {Nitta}},\ }\href
  {https://doi.org/10.1103/PhysRevResearch.2.013193} {\bibfield  {journal}
  {\bibinfo  {journal} {Phys. Rev. Research}\ }\textbf {\bibinfo {volume}
  {2}},\ \bibinfo {pages} {013193} (\bibinfo {year} {2020})}\BibitemShut
  {NoStop}%
\bibitem [{\citenamefont {Masaki}\ \emph {et~al.}(2021)\citenamefont {Masaki},
  \citenamefont {Mizushima},\ and\ \citenamefont {Nitta}}]{Masaki:2021hmk}%
  \BibitemOpen
  \bibfield  {author} {\bibinfo {author} {\bibfnamefont {Y.}~\bibnamefont
  {Masaki}}, \bibinfo {author} {\bibfnamefont {T.}~\bibnamefont {Mizushima}},\
  and\ \bibinfo {author} {\bibfnamefont {M.}~\bibnamefont {Nitta}},\
  }\href@noop {} {\  (\bibinfo {year} {2021})},\ \Eprint
  {https://arxiv.org/abs/2107.02448} {arXiv:2107.02448 [cond-mat.supr-con]}
  \BibitemShut {NoStop}%
\bibitem [{\citenamefont {Mizushima}\ \emph {et~al.}(2021)\citenamefont
  {Mizushima}, \citenamefont {Yasui}, \citenamefont {Inotani},\ and\
  \citenamefont {Nitta}}]{Mizushima:2021qrz}%
  \BibitemOpen
  \bibfield  {author} {\bibinfo {author} {\bibfnamefont {T.}~\bibnamefont
  {Mizushima}}, \bibinfo {author} {\bibfnamefont {S.}~\bibnamefont {Yasui}},
  \bibinfo {author} {\bibfnamefont {D.}~\bibnamefont {Inotani}},\ and\ \bibinfo
  {author} {\bibfnamefont {M.}~\bibnamefont {Nitta}},\ }\href
  {https://doi.org/10.1103/PhysRevC.104.045803} {\bibfield  {journal} {\bibinfo
   {journal} {Phys. Rev. C}\ }\textbf {\bibinfo {volume} {104}},\ \bibinfo
  {pages} {045803} (\bibinfo {year} {2021})},\ \Eprint
  {https://arxiv.org/abs/2108.01256} {arXiv:2108.01256 [nucl-th]} \BibitemShut
  {NoStop}%
\bibitem [{\citenamefont {Abud}\ and\ \citenamefont
  {Sartori}(1981)}]{Abud:1981tf}%
  \BibitemOpen
  \bibfield  {author} {\bibinfo {author} {\bibfnamefont {M.}~\bibnamefont
  {Abud}}\ and\ \bibinfo {author} {\bibfnamefont {G.}~\bibnamefont {Sartori}},\
  }\href {https://doi.org/10.1016/0370-2693(81)90578-5} {\bibfield  {journal}
  {\bibinfo  {journal} {Phys. Lett. B}\ }\textbf {\bibinfo {volume} {104}},\
  \bibinfo {pages} {147} (\bibinfo {year} {1981})}\BibitemShut {NoStop}%
\bibitem [{\citenamefont {Abud}\ and\ \citenamefont
  {Sartori}(1983)}]{Abud:1983id}%
  \BibitemOpen
  \bibfield  {author} {\bibinfo {author} {\bibfnamefont {M.}~\bibnamefont
  {Abud}}\ and\ \bibinfo {author} {\bibfnamefont {G.}~\bibnamefont {Sartori}},\
  }\href {https://doi.org/10.1016/0003-4916(83)90017-9} {\bibfield  {journal}
  {\bibinfo  {journal} {Annals Phys.}\ }\textbf {\bibinfo {volume} {150}},\
  \bibinfo {pages} {307} (\bibinfo {year} {1983})}\BibitemShut {NoStop}%
\bibitem [{\citenamefont {Nitta}(1999)}]{Nitta:1998qp}%
  \BibitemOpen
  \bibfield  {author} {\bibinfo {author} {\bibfnamefont {M.}~\bibnamefont
  {Nitta}},\ }\href {https://doi.org/10.1142/S0217751X99001202} {\bibfield
  {journal} {\bibinfo  {journal} {Int. J. Mod. Phys. A}\ }\textbf {\bibinfo
  {volume} {14}},\ \bibinfo {pages} {2397} (\bibinfo {year} {1999})},\ \Eprint
  {https://arxiv.org/abs/hep-th/9805038} {arXiv:hep-th/9805038} \BibitemShut
  {NoStop}%
\bibitem [{\citenamefont {Barnett}\ \emph {et~al.}(2006)\citenamefont
  {Barnett}, \citenamefont {Turner},\ and\ \citenamefont
  {Demler}}]{Barnett:PRL2006}%
  \BibitemOpen
  \bibfield  {author} {\bibinfo {author} {\bibfnamefont {R.}~\bibnamefont
  {Barnett}}, \bibinfo {author} {\bibfnamefont {A.}~\bibnamefont {Turner}},\
  and\ \bibinfo {author} {\bibfnamefont {E.}~\bibnamefont {Demler}},\ }\href
  {https://doi.org/10.1103/PhysRevLett.97.180412} {\bibfield  {journal}
  {\bibinfo  {journal} {Phys. Rev. Lett.}\ }\textbf {\bibinfo {volume} {97}},\
  \bibinfo {pages} {180412} (\bibinfo {year} {2006})}\BibitemShut {NoStop}%
\bibitem [{\citenamefont {Kawaguchi}\ and\ \citenamefont
  {Ueda}(2011)}]{Kawaguchi:PRA2011}%
  \BibitemOpen
  \bibfield  {author} {\bibinfo {author} {\bibfnamefont {Y.}~\bibnamefont
  {Kawaguchi}}\ and\ \bibinfo {author} {\bibfnamefont {M.}~\bibnamefont
  {Ueda}},\ }\href {https://doi.org/10.1103/PhysRevA.84.053616} {\bibfield
  {journal} {\bibinfo  {journal} {Phys. Rev. A}\ }\textbf {\bibinfo {volume}
  {84}},\ \bibinfo {pages} {053616} (\bibinfo {year} {2011})}\BibitemShut
  {NoStop}%
\bibitem [{\citenamefont {Lian}\ \emph {et~al.}(2012)\citenamefont {Lian},
  \citenamefont {Ho},\ and\ \citenamefont {Zhai}}]{Lian:PRA2012}%
  \BibitemOpen
  \bibfield  {author} {\bibinfo {author} {\bibfnamefont {B.}~\bibnamefont
  {Lian}}, \bibinfo {author} {\bibfnamefont {T.-L.}\ \bibnamefont {Ho}},\ and\
  \bibinfo {author} {\bibfnamefont {H.}~\bibnamefont {Zhai}},\ }\href
  {https://doi.org/10.1103/PhysRevA.85.051606} {\bibfield  {journal} {\bibinfo
  {journal} {Phys. Rev. A}\ }\textbf {\bibinfo {volume} {85}},\ \bibinfo
  {pages} {051606(R)} (\bibinfo {year} {2012})}\BibitemShut {NoStop}%
\bibitem [{\citenamefont {Stamper-Kurn}\ and\ \citenamefont
  {Ueda}(2013)}]{Stamper-Kurn:APS2013}%
  \BibitemOpen
  \bibfield  {author} {\bibinfo {author} {\bibfnamefont {D.~M.}\ \bibnamefont
  {Stamper-Kurn}}\ and\ \bibinfo {author} {\bibfnamefont {M.}~\bibnamefont
  {Ueda}},\ }\href {https://doi.org/10.1103/RevModPhys.85.1191} {\bibfield
  {journal} {\bibinfo  {journal} {Rev. Mod. Phys.}\ }\textbf {\bibinfo {volume}
  {85}},\ \bibinfo {pages} {1191} (\bibinfo {year} {2013})}\BibitemShut
  {NoStop}%
\bibitem [{\citenamefont {Takahashi}(2015)}]{Takahashi:2015}%
  \BibitemOpen
  \bibfield  {author} {\bibinfo {author} {\bibfnamefont {D.~A.}\ \bibnamefont
  {Takahashi}},\ }\href {https://doi.org/10.7566/JPSJ.84.025001} {\bibfield
  {journal} {\bibinfo  {journal} {Journal of the Physical Society of Japan}\
  }\textbf {\bibinfo {volume} {84}},\ \bibinfo {pages} {025001} (\bibinfo
  {year} {2015})},\ \Eprint
  {https://arxiv.org/abs/https://doi.org/10.7566/JPSJ.84.025001}
  {https://doi.org/10.7566/JPSJ.84.025001} \BibitemShut {NoStop}%
\bibitem [{\citenamefont {Mizushima}\ and\ \citenamefont
  {Nitta}(2018)}]{Mizushima:2017pma}%
  \BibitemOpen
  \bibfield  {author} {\bibinfo {author} {\bibfnamefont {T.}~\bibnamefont
  {Mizushima}}\ and\ \bibinfo {author} {\bibfnamefont {M.}~\bibnamefont
  {Nitta}},\ }\href {https://doi.org/10.1103/PhysRevB.97.024506} {\bibfield
  {journal} {\bibinfo  {journal} {Phys. Rev.}\ }\textbf {\bibinfo {volume}
  {B97}},\ \bibinfo {pages} {024506} (\bibinfo {year} {2018})},\ \Eprint
  {https://arxiv.org/abs/1710.07403} {arXiv:1710.07403 [cond-mat.supr-con]}
  \BibitemShut {NoStop}%
%%CITATION = ARXIV:1710.07403;%%
\end{thebibliography}%

\end{document}